\documentclass[12pt]{iopart}
\usepackage[french,english]{babel}
\usepackage{graphicx}
\usepackage{amsthm}
\expandafter\let\csname equation*\endcsname\relax
\expandafter\let\csname endequation*\endcsname\relax
\usepackage{amsmath}
\usepackage{amsfonts}
\usepackage{amssymb}
\usepackage{longtable}
\usepackage{afterpage}
\usepackage{textcomp}
\usepackage[normalem]{ulem}

\newcommand{\bv}[1]{\mathbf{#1}}

\begin{document}

\title{Nonlinear saturation of resistive tearing modes in a cylindrical tokamak with and without solving the dynamics}

\author{J. Loizu$^1$ and D. Bonfiglio$^{2,3}$}
\address{$^1$ \'Ecole Polytechnique F\'ed\'erale de Lausanne, Swiss Plasma Center, CH-1015 Lausanne, Switzerland \\
 $^2$  Consorzio RFX, CNR, ENEA, INFN, Universit\`a di Padova, Acciaierie Venete SpA, Corso Stati Uniti 4, 35127 Padova, Italy \\
 $^3$ CNR-ISTP, Padova, Italy}
\ead{joaquim.loizu@epfl.ch}
\begin{abstract}
We show that the saturation of resistive tearing modes in a cylindrical tokamak, as well as the corresponding island width, can be directly calculated with an MHD equilibrium code without solving the dynamics and without considering resistivity. The results are compared to initial value resistive MHD simulations and to an analytical nonlinear theory. For small enough islands, the agreement is remarkable. For sufficiently large islands, the equilibrium calculations, which assume a flat current profile inside the island, overestimate the saturation amplitude. On the other hand, excellent agreement between nonlinear resistive MHD simulations and nonlinear theory is observed for all the considered tearing unstable equilibria.
\end{abstract}

\maketitle

\section{Introduction} \label{sec:intro}

In magnetic fusion, tearing modes are an important class of instabilities that can often be described by resistive magnetohydrodynamics (MHD) \cite{White2014}. Tearing modes are current-driven internal instabilities that drive spontaneous magnetic reconnection at rational surfaces in toroidally confined plasmas. Magnetic islands grow in time, thus potentially affecting confinement, and eventually saturate nonlinearly. If their saturated amplitude is too large, they can trigger disruptions, sometimes even inducing the complete termination of the plasma discharge \cite{LaHaye2006}. While resistivity is required to allow these modes to grow, their saturated states are 3D MHD equilibria that are often independent of resistivity, at least for sufficiently large Lundquist numbers \cite{Poye2014}. This suggests that a direct MHD equilibrium computation might be able to describe and even predict these saturated states, as was shown to be the case for the saturation of ideal instabilities in tokamaks \cite{Cooper2010, Cooper2011, Kleiner2019} or the formation of saturated double-axis states in reversed field pinches \cite{Dennis2013a}. The possibility of finding these nonlinearly saturated tearing modes directly, namely without solving the dynamics and without free parameters, was investigated and successfully shown for the first time in slab geometry \cite{Loizu2020}. More recently, a similar type of calculation was carried out in cylindrical geometry for the case of externally forced reconnection in a cylinder \cite{Wright2022}, although the volume available for reconnection remained unconstrained. Here we investigate the possibility of directly predicting the saturated state of tearing modes in a cylindrical tokamak geometry at zero pressure. In this configuration, we can exploit an exact nonlinear saturation theory \cite{Arcis2006} that predicts the saturated island width for a given initially axisymmetric equilibrium that is unstable to a tearing mode. The results are compared to those obtained with the SpeCyl code \cite{Cappello1996,Bonfiglio2010}, which solves the time-dependent, nonlinear visco-resistive MHD equations in a cylinder. These predictions serve as a benchmark for the direct equilibrium approach, where we use the Stepped-Pressure Equilibrium Code (SPEC) \cite{Hudson2012, Hudson2020} to describe, without free parameters, the spontaneous formation of these nonlinearly saturated tearing islands. In general, the direct calculation of saturated states with an equilibrium approach is interesting for two reasons: on the one hand, it provides insights into the nature of the saturated state itself; on the other hand, the approach is potentially much faster than the initial value approach, especially as we consider larger systems (such as magnetic fusion reactors) or more complex geometries (such as stellarators), where analytical theories are not available and initial value approaches are very demanding.

The remainder of this paper is organized as follows. In Sec.~\ref{sec:eq} we construct and analyze the linear stability of a family of MHD equilibria both analytically and with SPEC. Section \ref{sec:nltheory} describes a nonlinear theory that predicts the saturated island width of these unstable equilibria. In Sec.~\ref{sec:nlsim} we carry out nonlinear simulations with the SpeCyl code and compare the obtained saturated tearing modes with the theoretical predictions. Section \ref{sec:nlspec} describes how SPEC can also be used to retrieve the saturated states corresponding to each unstable equilibrium, and a discussion on the level of agreement is provided. Conclusions follow in Sec.~\ref{sec:conc}.
 
\section{Family of equilibria and linear stability analysis} \label{sec:eq}

\begin{figure}[h!]
\centering
\includegraphics[clip=true, scale=0.52]{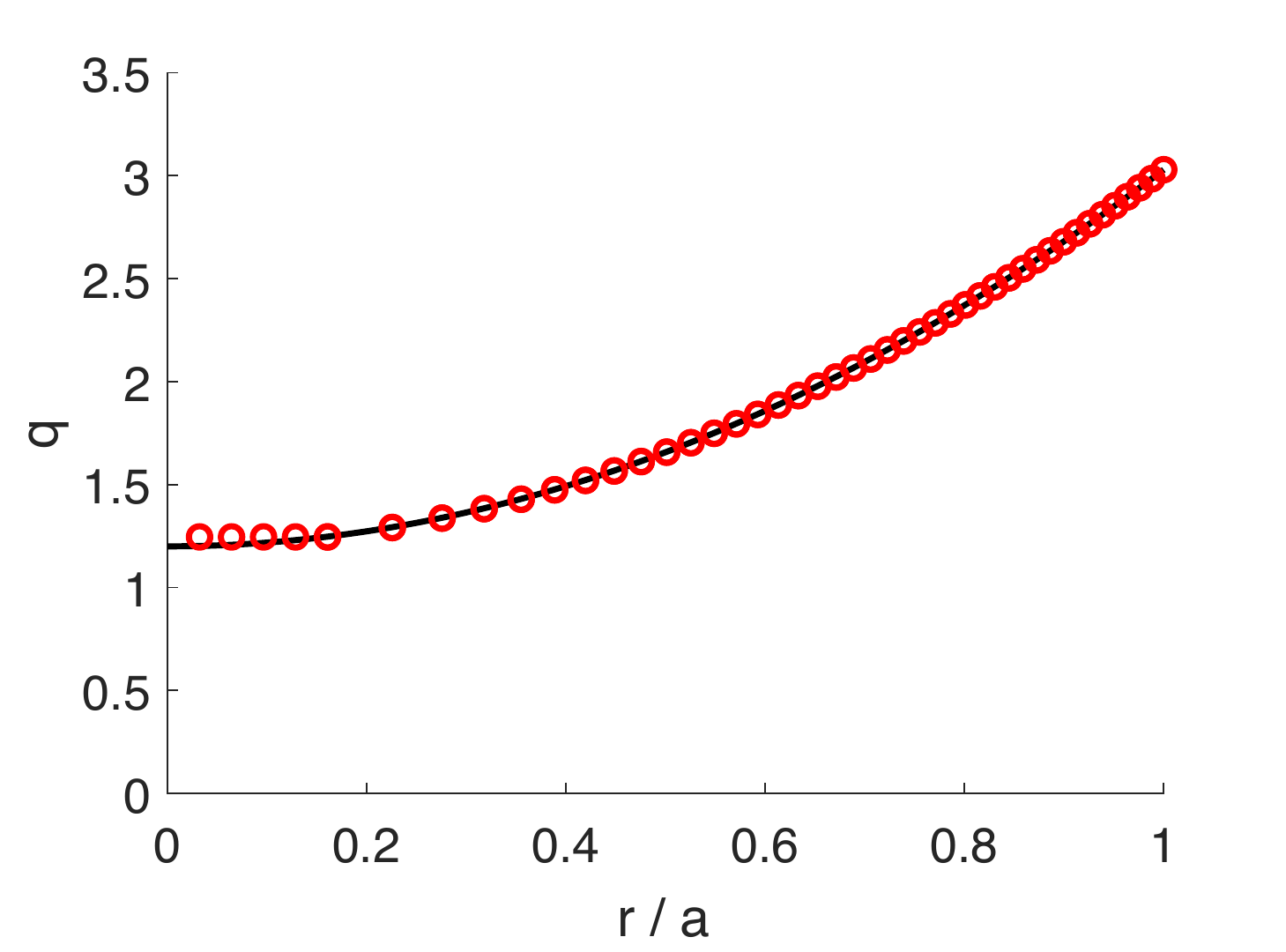} 
\includegraphics[clip=true, scale=0.52]{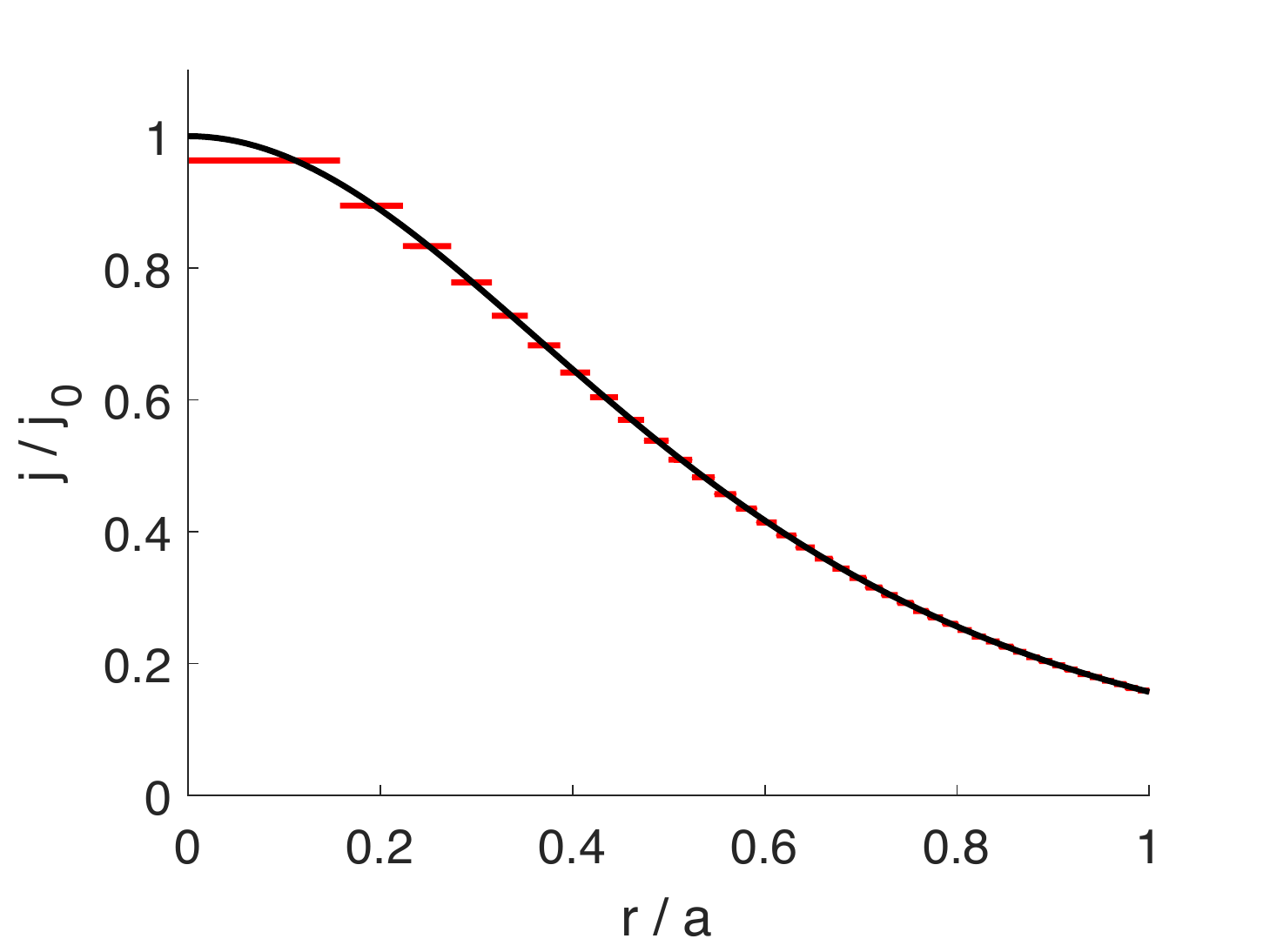}
\caption{Equilibrium profiles for the safety factor (left) and current density (right) as given by Eqs.~(\ref{qprof})-(\ref{jprof}) (solid black lines) and as obtained from SPEC (red circles and lines) with $N_v=41$. Here $q_0=1.2$ and $r_0/a=0.81$.} \label{fig:eqprof} 
\end{figure}

We construct a family of non-rotating, cylindrical MHD equilibria with zero pressure and parametrized by the on-axis safety factor $q_0$:

\begin{equation}
q(r) = q_0\left[1+\left(\frac{r}{r_0}\right)^2\right] \label{qprof}
\end{equation}
for $0\leq r\leq a$ and where $a$ and $r_0$ are fixed and correspond, respectively, to the radius of the plasma boundary and to the current channel width. As we will see, the value of $q_0$ controls the stability of the equilibrium, the presence and location of resonant rational surfaces, as well as the width of the saturated islands. In the strong guide field limit ($B_z\gg B_{\theta}$), which for a tokamak with $q\sim 1$ also corresponds to the large aspect ratio limit ($R\gg a$), the axial current density profile associated to this $q$-profile is given by:
\begin{equation}
j_z(r) = \frac{j_0}{\left[1+\left(\frac{r}{r_0}\right)^2\right]^2} \label{jprof}
\end{equation}
where $j_0=2B_0 / (\mu_0q_0 R)$ is the on-axis current density, $B_0=B_z(0)$, and $2\pi R$ is the length of the cylinder, which is considered to be $2\pi$-periodic in both $\theta$ and $\varphi=z/R$. Figure \ref{fig:eqprof} shows examples of profiles corresponding to Eqs.~(\ref{qprof})-(\ref{jprof}). Given the fixed, circular plasma boundary at $r=a$, and given $q_0$, $r_0$, $B_0$, and $R$, the ideal MHD equilibrium magnetic field $\bv{B}$ is fully determined by the force-balance equation $\bv{j}\times\bv{B}=\nabla p=0$, which reduces to a one-dimensional, first-order differential equation for $B_z(r)$. These equilibria are ideally stable as long as $q_0>1$ \cite{Freidberg2014}, but as we will see in the next section, they can be resistive unstable \cite{Furth1973}. \\
We can also construct these equilibria numerically by using the Stepped-Pressure Equilibrium Code (SPEC) \cite{Hudson2012, Hudson2020}, with the purpose of seeking nearby, relaxed (lower potential energy) equilibria, which can be accessed via the possible opening of magnetic islands. SPEC considers a finite number $N_v$ of nested volumes, hereafter referred to as \emph{relaxation volumes}, that are separated by magnetic surfaces, hereafter referred to as \emph{ideal interfaces}. In cylindrical geometry, the position of each ideal interface $l$ is generally described by $\bv{x}_l(\theta,\varphi)=r_l(\theta,\varphi)\cos{\theta}\ \hat{i} + r_l(\theta,\varphi)\sin{\theta}\ \hat{j} + R\varphi \hat{k}$ and thus can be determined by a set of Fourier coefficients, since we can write $r_l(\theta,\varphi)=\sum_{m,n} r_{l,mn}\cos{(m\theta-n\varphi)}$. In each relaxation volume, the magnetic field solves a Beltrami equation $\nabla\times\bv{B}=\mu \bv{B}$ with $\mu$ a constant that is related to the parallel current density. The solution in the volumes is then determined by the shape of the boundary interfaces where $\bv{B}\cdot\bv{n}=0$, the value of $\mu$, and two scalars $\Psi_p$ and $\Psi_t$ that correspond to the "poloidal" (constant-$\theta$) and "toroidal" (constant-$\varphi$) magnetic fluxes enclosed by each volume (in the innermost volume, only $\Psi_t$ is required). The position and shape of each interface is unknown \emph{a priori} and is determined by a jump condition across each interface, $[[B^2]]=B^2_+ - B^2_- = 0$, where $+$ and $-$ refers to each side of the interface. This condition is the local equivalent of force balance and ensures that the inner and outer pressures are in balance everywhere on an interface. We remark that equilibria with finite plasma pressure can also be described in SPEC but here we are considering zero-pressure equilibria. The magnetic fields satisfying these equations are extrema of the plasma potential energy, $W=\int_{V_p} B^2 / (2\mu_0) dV$, when considering arbitrary variations in the magnetic field inside the relaxation volumes and ideal variations of the interface geometries, with the constraint of a conserved magnetic helicity, $K_l=\int_{V_l} \bv{A}\cdot\bv{B} \ dV$, and magnetic fluxes in each relaxation volume \cite{Hudson2012}. Here $V_p=\sum_{l=1}^{N_v} V_l$ is the total plasma volume and $\bv{A}$ is the magnetic vector potential. Furthermore, it has been shown that the magnetic field satisfying these equations retrieves exactly ideal MHD in the limit $N_v\rightarrow \infty$ \cite{Dennis2013}. Hence SPEC can be used to construct the family of ideal MHD equilibria described by Eq. (\ref{qprof}) if a large value of $N_v$ is chosen and if the appropriate constraints are provided. In fact, the constants ($\mu$, $\Psi_p$, $\Psi_t$) in each volume, which determine the Beltrami fields, can be related to other physical quantities such as the safety factor $q$ on each side of each interface, $q_+$ and $q_-$, or the net-toroidal-current flowing in each volume, $I_{\mathrm{vol}}$,  and on each interface, $I_{\mathrm{surf}}$ \cite{Baillod2021}. One can thus for example constrain in each volume $(q_+, q_-, \Psi_t)$ if the safety factor is to be imposed, or $(I_{\mathrm{vol}} , I_{\mathrm{surf}}, \Psi_t)$ if the current profile is to be imposed. Figure \ref{fig:eqprof} shows an example of an equilibrium reconstructed with SPEC, where $N_v=41$ and the safety factor was constrained on each interface by following Eq. (\ref{qprof}). After the SPEC calculation, one can look at the global behaviour of $q(r)$ as well as that of $j_z(r)$, which indeed reproduce well the analytical profiles.

\subsection{Linear stability from classical tearing theory} 

The linear resistive stability of a cylindrical tokamak with zero pressure is uniquely determined by the sign of $\Delta^\prime$ \cite{Furth1973}, 

\begin{equation}
\Delta^\prime = \lim_{\epsilon\rightarrow 0^+}\frac{\hat{\psi}'(r_s+\epsilon)-\hat{\psi}'(r_s-\epsilon)}{\hat{\psi}(r_s)} \label{dp}
\end{equation}
with $\Delta^\prime>0$ implying instability. Here $\hat{\psi}(r)$ is the amplitude of the perturbed flux function $\psi(r,\theta,\varphi)=\hat{\psi}(r)\cos{(m\theta-n\varphi)}$ for the linear ideal mode $(m,n)$ associated to the rational surface $q(r_s)=q_s=m/n$, where $m$ and $n$ are respectively the poloidal (along $\theta$) and toroidal (along $\varphi$) mode numbers. The corresponding magnetic field perturbation is $\delta \bv{B}=\nabla\psi\times\bv{\hat{z}}$, and thus the radial magnetic field perturbation, $\delta B_r$, for this mode is proportional to $\hat{\psi}$. Notice that $\Delta^\prime$ depends on equilibrium profiles but is independent of resistivity. It can be calculated from the linearized ideal MHD equations, by matching the so-called ``outer solutions". A shooting code provides a numerical solution for $\hat{\psi}(r)$ and a value of $\Delta^\prime$ for a given resonant surface. For the family of equilibria described by Eq.~(\ref{qprof}), choosing $r_0/a=0.81$, $R/a=10$, and assuming the presence of a perfectly conducting shell at the plasma boundary, we find instability for the $m=2$, $n=1$ mode whenever $1<q_0<2$. In that range of $q_0$ values, we also find no other instability, hence we can focus on the $q_s=2$ resonant surface. Figure \ref{fig:psi} shows an example of solution for $\hat{\psi}(r)$ for $q_0=1.2$. Figure \ref{fig:dpp} shows the obtained values of $\Delta^\prime$ as a function of $q_0$. For larger values of $q_0$, the value of $\Delta^\prime$ keeps increasing but of course for $q_0>2$  there is no instability because there is no $q_s=2$ surface anymore.

\subsection{Linear stability from SPEC} \label{sec:specstab}

\begin{figure}[h!]
\centering
\includegraphics[clip=true, scale=0.8]{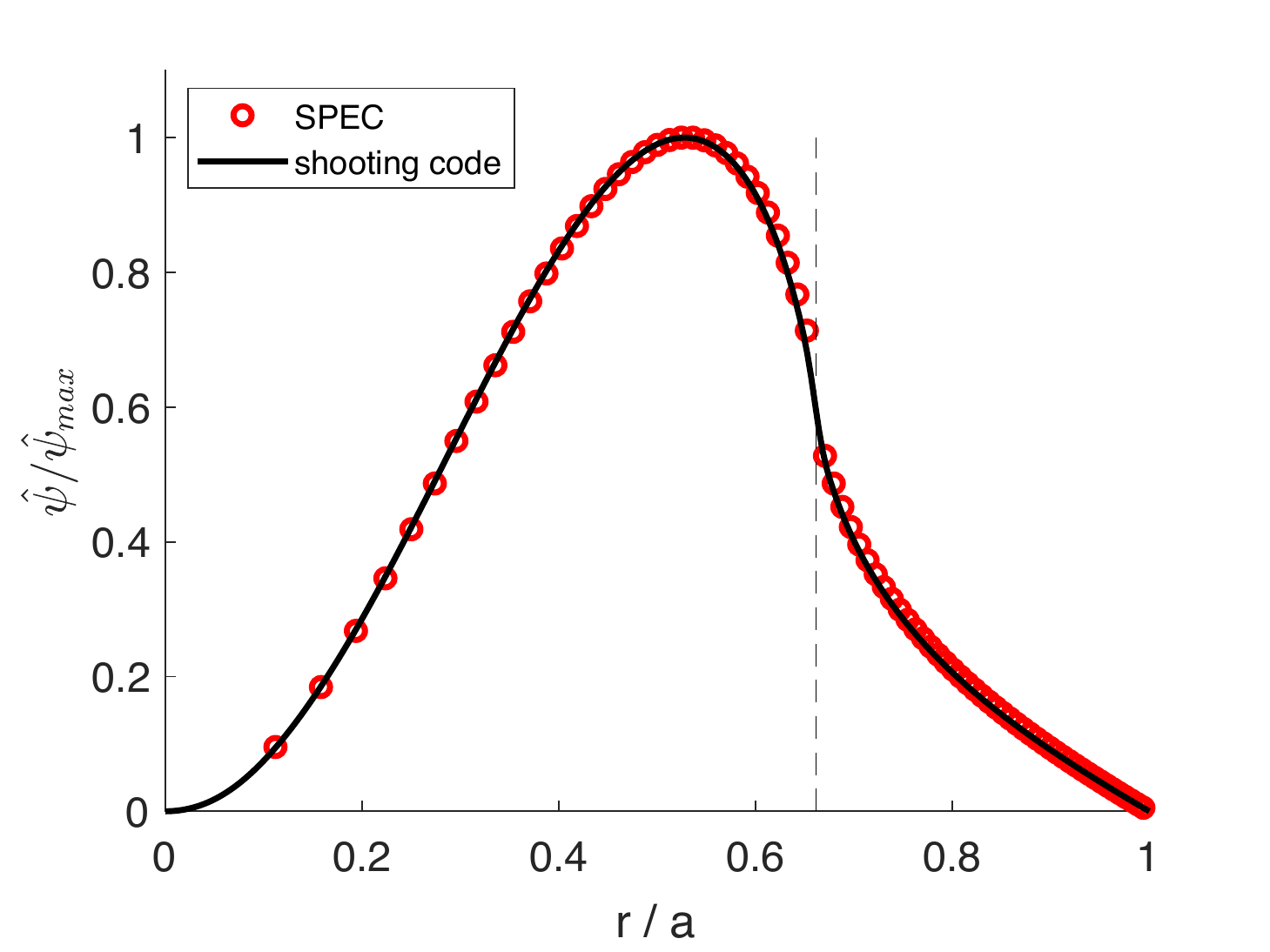}
\caption{Radial profile of the perturbed flux function $\hat{\psi}$ for the unstable $m=2,n=1$ mode obtained from the linearized ideal MHD equations (black solid line). The same function can be inferred from the linear radial displacement $\xi(r)$ obtained with SPEC using the relation $\hat{\psi}(r)=rB_{z0}(1/q(r)-1/q_s)\xi(r)/R$. The initial equilibrium is the same as for Figure \ref{fig:eqprof}. The dashed line indicates $r=r_s$.} \label{fig:psi} 
\end{figure}

\begin{figure}[h!]
\centering
\includegraphics[clip=true, scale=0.75]{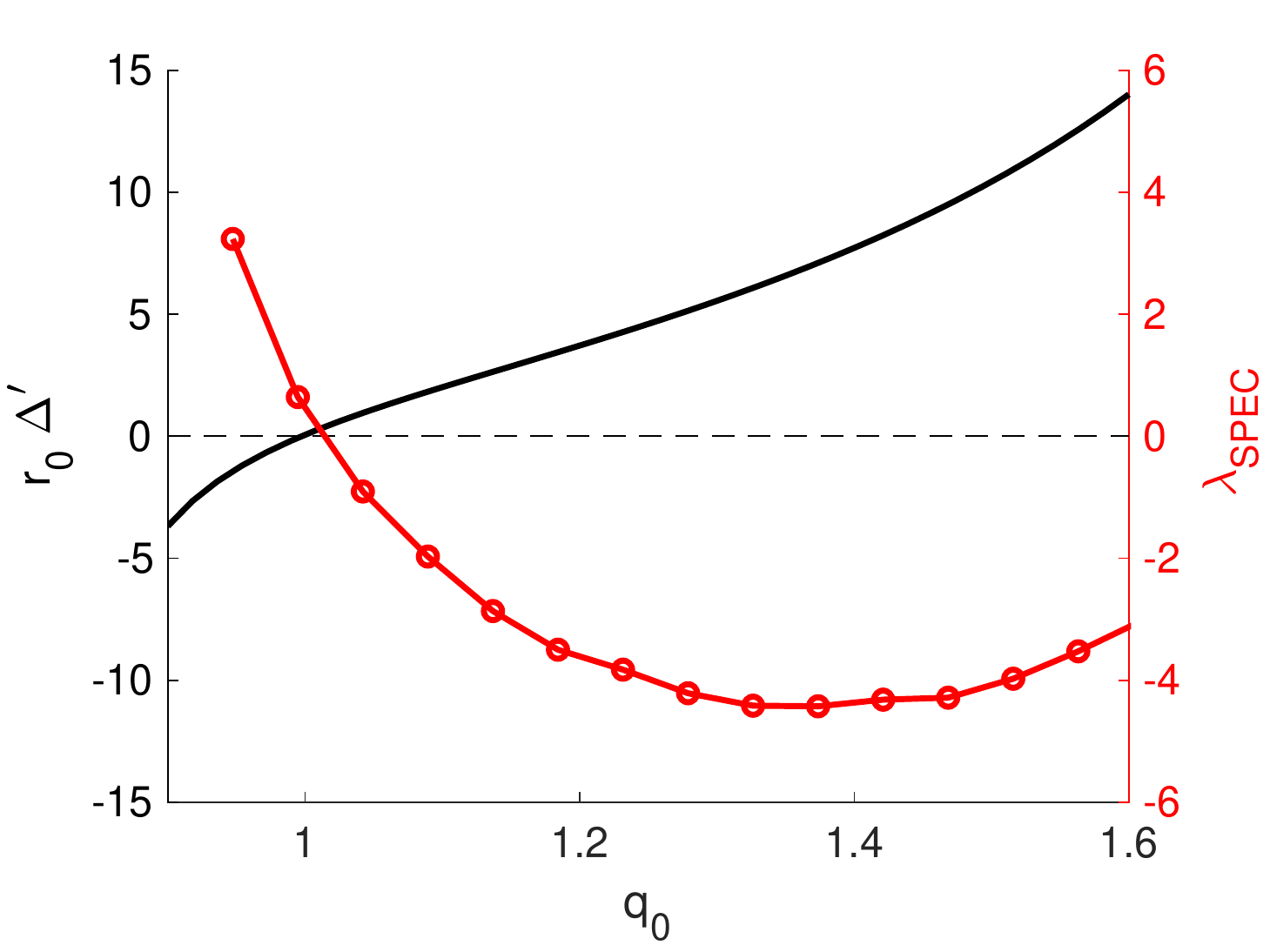}
\caption{$\Delta^\prime$ versus $q_0$ as obtained from a shooting code (black solid line) for the $m=2,n=1$ mode. The equilibrium $q$ profile is given in Eq.~(\ref{qprof}) with $r_0/a=0.81$, and $R/a=10$. The smallest eigenvalue of the force-gradient matrix from SPEC, $\lambda_{SPEC}$, is also shown in arbitrary units for each value of $q_0$ (red circles).} \label{fig:dpp} 
\end{figure}

The stability of each of these equilibria can also be assessed with the SPEC code by evaluating the eigenvalues of the Hessian matrix $\mathcal{H}$, whose coefficients measure the second variation of the plasma potential energy with respect to perturbations in the geometry of the ideal interfaces, with the constraint of fixed magnetic helicity and fluxes in each volume  \cite{Loizu2019a, Kumar2021, Kumar2022}. In cylindrical geometry, however, one can also directly evaluate the eigenvalues of the force-gradient matrix $\mathcal{G}$, whose coefficients measure the variation in the force on each interface, $f_l = [[B^2/2\mu_0]]_l$, with respect to perturbations in the geometry of the ideal interfaces, $r_{l,mn}$, with the constraint of fixed magnetic helicity and fluxes in each volume (see Appendix A for more details). Let us call $\lambda$ the smallest eigenvalue of $\mathcal{G}$. If $\lambda \geq0$ the equilibrium is stable. If $\lambda<0$, then the equilibrium is unstable and the eigenmode $\bv{u}_{\lambda}\equiv\{r_{l,mn}\}$ corresponding to the eigenvalue $\lambda$ provides information about the mode number $(m,n)$ of the instability as well as its radial structure. For the family of equilibria described by Eq.~(\ref{qprof}), we find that for $1<q_0<2$ there is a negative eigenvalue (thus instability) that corresponds to the $m=2$, $n=1$ mode (Figure \ref{fig:dpp}). Outside this interval of $q_0$-values, all eigenvalues are positive (thus stability). In the case of instability, the radial structure of the unstable eigenmode agrees well with the one obtained from the shooting code, as shown in Fig. \ref{fig:psi}. Indeed, this comparison is carried out by relating the radial displacement of the flux surfaces to the perturbed flux function, $\delta \bv{B} = \nabla \times ( \xi \times \bv{B}_0) = \nabla \psi \times \bv{\hat{z}}$, and evaluating the radial displacement of the SPEC ideal interfaces, which has an amplitude $\xi(r_l)=r_{l,mn}-r_{l,00}$ for each mode $(m,n)$.

\section{Nonlinear saturation theory} \label{sec:nltheory}

\begin{figure}[h!]
\centering
\includegraphics[viewport=0 200 550 650, clip=true, scale=0.65]{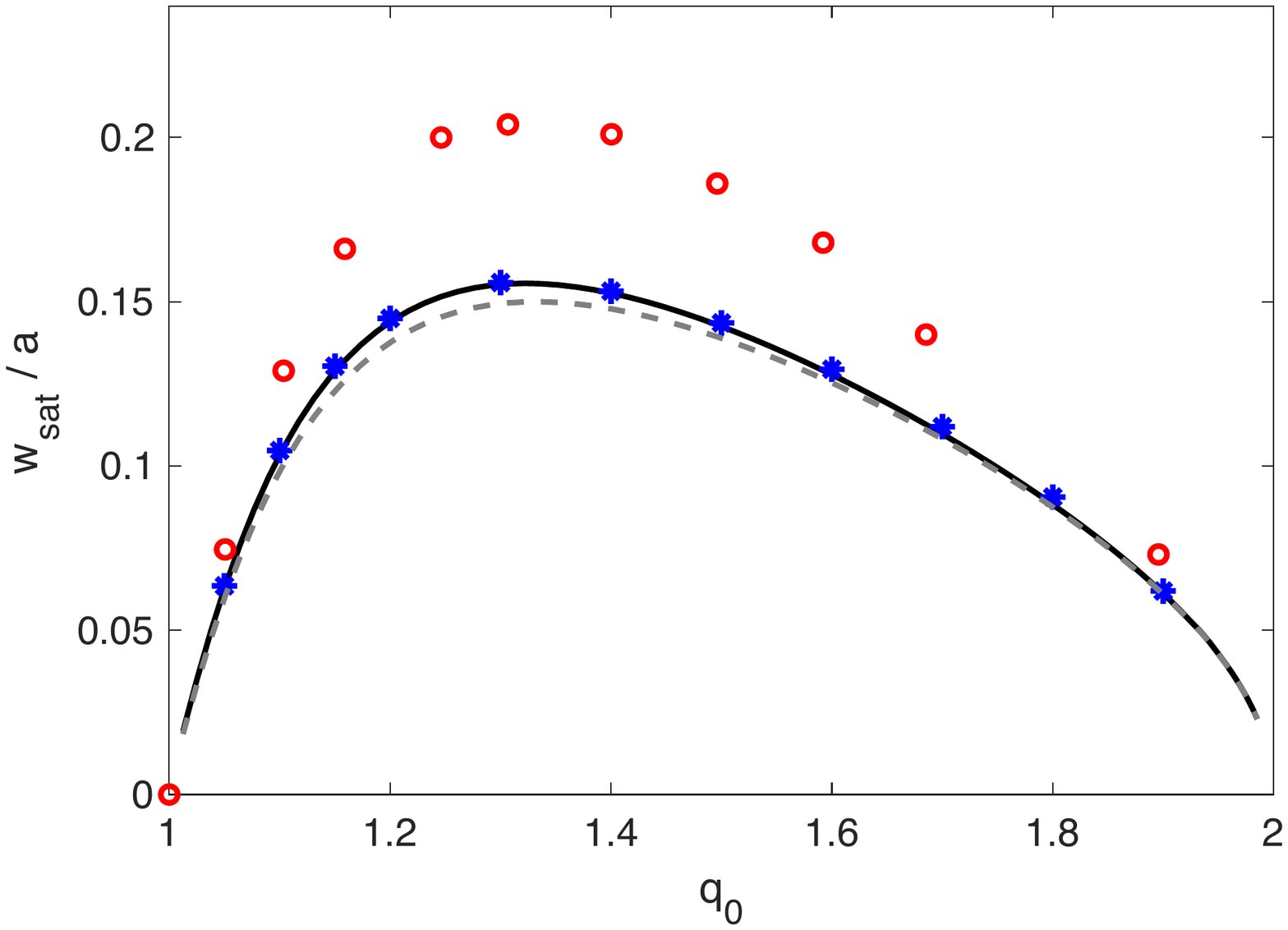}
\caption{$w_{sat}$ versus $q_0$ as obtained from Eq.~(\ref{wsat}) with $\sigma=1$ (solid black line) and $\sigma=0$ (dashed grey line). Also shown are the predictions from the initial-value code SpeCyl (blue stars) and from the equilibrium code SPEC (red circles). The initially unstable equilibria are the same as for Fig. \ref{fig:dpp}.} \label{fig:wsat} 
\end{figure}

The growth of the tearing mode promotes the growth of a magnetic island. The rate $\gamma$ at which the island grows depends on the plasma resistivity $\eta$, both in the linear \cite{Furth1973} and nonlinear \cite{Rutherford1973} phases. The saturation of the tearing mode, which in itself proves the existence of a nearby, generally three-dimensional MHD equilibrium, can be characterized by the width of the saturated island, $w_{sat}$, and that value becomes independent of resistivity for sufficiently small resistivity \cite{Poye2014}. For a cylindrical tokamak, Arcis et al \cite{Arcis2006} derived a nonlinear theory for the time evolution of the island width $w(t)$,
\begin{equation}
\hspace*{-2cm}
\frac{\mu_0}{\eta_{eq}(r_s)}\frac{dw}{dt} = 1.22\Delta^\prime + w\left\{ \frac{\mathcal{A}^2}{2}\ln{\frac{w}{w_0} - 2.21\mathcal{A}^2 + 0.40\frac{\mathcal{A}}{r_s} + \frac{\mathcal{B}}{2}} + 0.17\sigma\frac{\mathcal{A}^2s}{2-s}\right\}  \label{wsat}
\end{equation}
that is exact to first order in $w$. The coefficients in Eq.~(\ref{wsat}) can all be calculated from the equilibrium profiles, namely, 
\begin{equation}
\hspace*{-1cm}
\mathcal{A} = \frac{j'_{eq}(r_s)}{j_{eq}(r_s)}\left(1-\frac{2}{s}\right)  \textrm{ ; } \\ 
\mathcal{B} = \frac{j''_{eq}(r_s)}{j_{eq}(r_s)}\left(1-\frac{2}{s}\right) \textrm{ ; } \\
w_0 = a \exp{(-\frac{\Sigma'}{2\mathcal{A}})} \label{arcispar}
\end{equation}
where $j_{eq}(r)$ is the equilibrium current density profile, which in our case is given by Eq.~(\ref{jprof}), $s = r_sq'(r_s)/q(r_s)$ is the magnetic shear at the resonant surface, 
\begin{equation}
\Sigma' = \lim_{\epsilon\rightarrow 0^+}\left[ \frac{\hat{\psi}'(r_s+\epsilon)+\hat{\psi}'(r_s-\epsilon)}{\hat{\psi}(r_s)} - 2\mathcal{A}\left(1+\ln{\left(\frac{\epsilon}{a}\right)}\right) \right] \label{sp}
\end{equation}
is a quantity that, similar to $\Delta^\prime$, can be obtained from the ideal outer solution, and $\sigma$ is a constant that defines whether the equilibrium resistivity is uniform ($\sigma=0$) or the equilibrium electric field is uniform ($\sigma=1$). Equation (\ref{wsat}) can be used to predict the saturated island width, $w_{sat}$, as a function only of the initial equilibrium profiles, by setting the left hand side to zero and solving for $w$ (notice that, as expected, the solution will not depend on $\eta$). This involves solving a single nonlinear algebraic equation and that can be easily done numerically. Figure \ref{fig:wsat} shows the predicted values of $w_{sat}$ for the same family of equilibria as in Sec.~\ref{sec:eq}, namely for different values of $q_0$. We observe that $w_{sat}$ increases with $q_0$ up to a maximum, $w_{sat}/a \approx 0.15$ at $q_0\approx1.3$, and then decreases again towards zero as $q_0$ approaches $q_s=2$. It is worth noticing that the value of $\Delta^\prime$, which gives a measure of the available energy for reconnection \cite{Furth1973}, is monotonically increasing with $q_0$ and thus the final amplitude of the mode is not entirely reflected by the amplitude of $\Delta^\prime$. We also observe that the choice of $\sigma$ (0 or 1) only introduces minor changes.

\section{Nonlinear resistive MHD simulations} \label{sec:nlsim}

The dynamics of the tearing mode, from the initially unstable equilibrium up until the nonlinear saturation of the island, can be calculated by using an \emph{initial-value} approach. Here we use the SpeCyl code \cite{Cappello2004,Bonfiglio2010} to solve the nonlinear visco-resistive MHD equations in a cylinder, in the constant-pressure, constant-density approximation. The model equations are, in dimensionless units,
\begin{eqnarray}
\frac{\partial \bv{B}}{\partial t} &=& -\nabla\times\bv{E} \nonumber \\
\bv{E} &=&  \eta \bv{j} - \bv{v}\times\bv{B} \nonumber \\
\bv{j} &=& \nabla\times\bv{B} \nonumber \\
\rho\frac{d \bv{v}}{d t} &=& \bv{j}\times\bv{B} + \rho\nu\nabla^2\bv{v} \nonumber \\
\nabla\cdot\bv{B}&=&0
\end{eqnarray}
where $\bv{E}$ and $\bv{B}$ are the electric and magnetic fields, $\bv{j}$ is the current density, $\bv{v}$ is the plasma velocity, $\frac{d \bv{v}}{d t}=\left[ \frac{\partial \bv{v}}{\partial t} + (\bv{v}\cdot\nabla)\bv{v} \right]$, and $\rho=m_in_0$ is the (constant) mass density. Lengths are normalized to the minor radius $a$, the plasma particle density to the uniform density $n_0$, the magnetic field to the initial value $B_0$ of the axial magnetic field on axis, the velocity to the Alfv\'en velocity $v_A=B_0/(\mu_0m_i n_0)^{1/2}$ and time to the Alfv\'en time $\tau_A=a/v_A$. In these units, the resistivity $\eta$ corresponds to the inverse Lundquist number $S^{-1}=\tau_A/\tau_{\eta}$, and the viscosity $\nu$ to the inverse viscous Lundquist number (for a scalar kinematic viscosity), $M^{-1}=\tau_A/\tau_{\nu}$, where $\tau_{\eta}$ and $\tau_{\nu}$ are the resistive time scale and viscous time scale, respectively. The equations are solved in cylindrical geometry $(r,\theta,z)$ and periodic boundary conditions are used for both the azimuthal coordinate $\theta$ and the axial coordinate $z$, which has periodicity $2\pi R$. The code uses finite differences in the radial coordinate $r$ and a spectral formulation in the two periodic coordinates $\theta$ and $z$. In the calculations presented in this paper, the Fourier harmonics $(m,n)=(2j,j)$ are retained, with $j=0,..,16$. Such large number of helical harmonics is required to properly capture the sharp flow patterns developing around the island separatrix, in particular for high Lundquist numbers. The boundary conditions at the plasma wall, $r=a$, are $B_r=E_{\theta}=0$ (perfect conductor), $E_z=E_0$ (externally imposed, constant, toroidal electric field), and $v_{\theta}=v_z=0$ (no slip boundary conditions). The rest of the boundary conditions can be derived from Ohm's law. More details about the SpeCyl code and its numerical algorithm can be found in Refs.~\cite{Cappello2004,Cappello1996}.

\begin{figure}[h!]
\centering
\includegraphics[clip=true, scale=0.52]{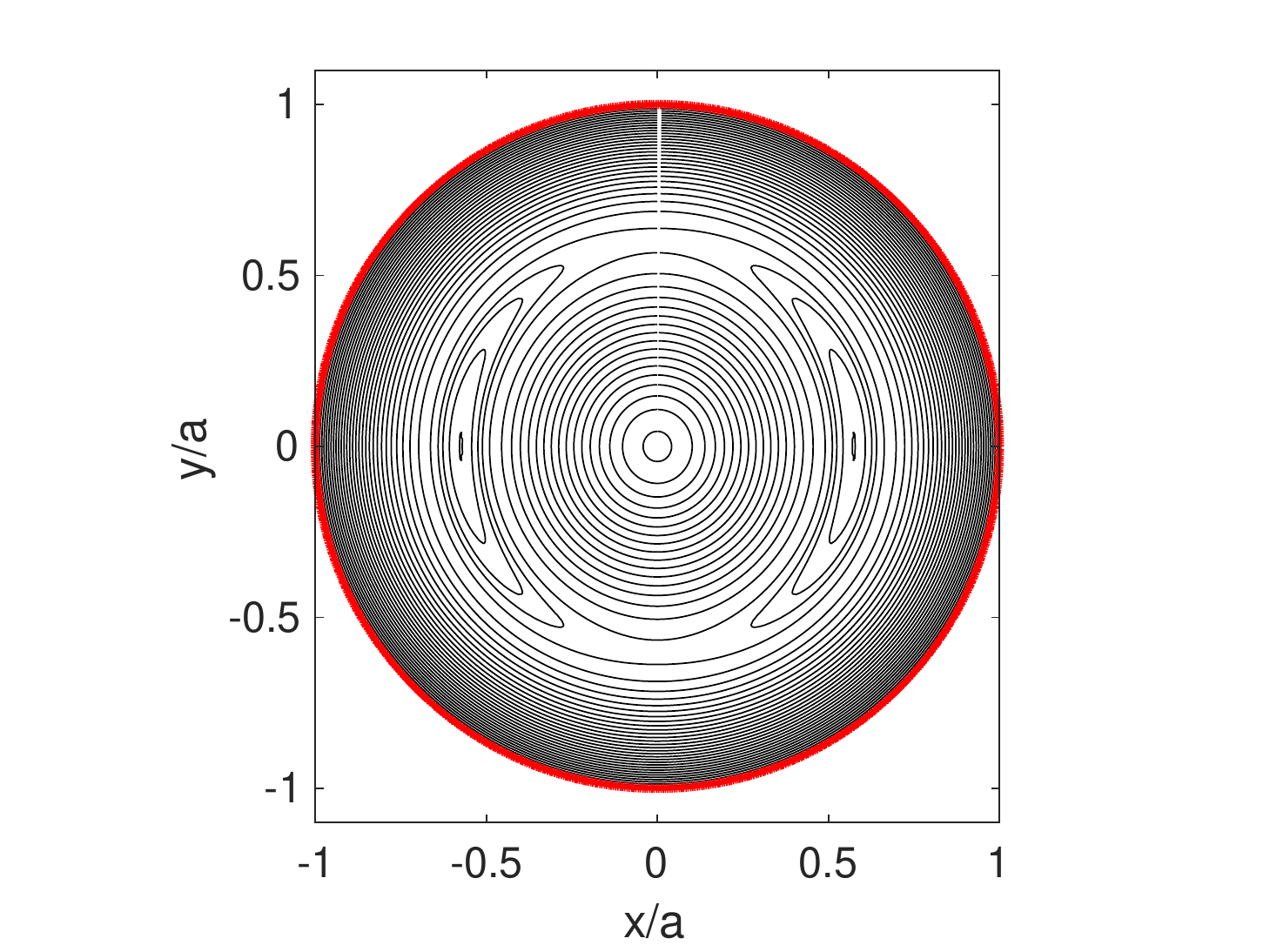} 
\includegraphics[clip=true, scale=0.52]{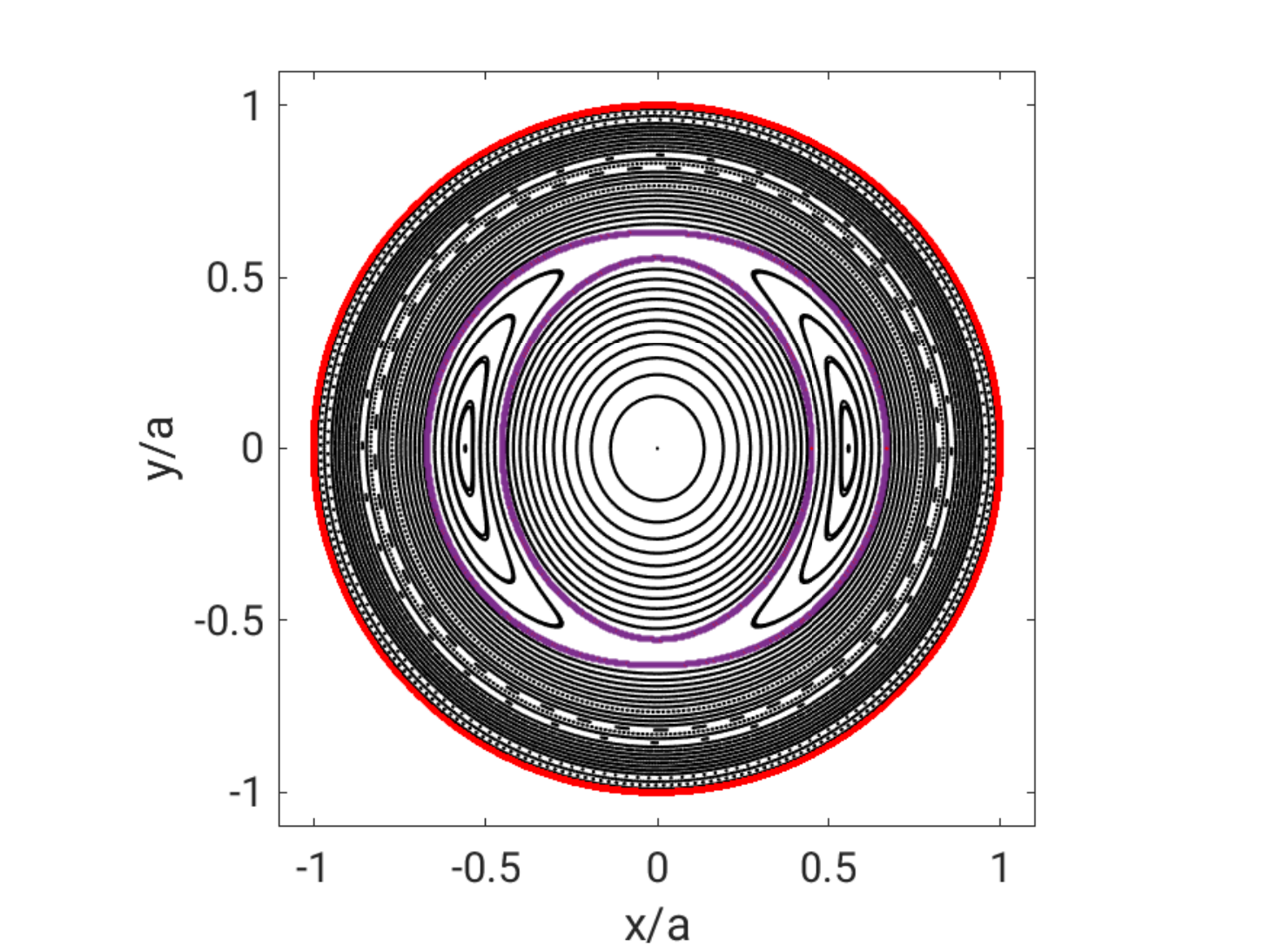}
\includegraphics[clip=true, scale=0.52]{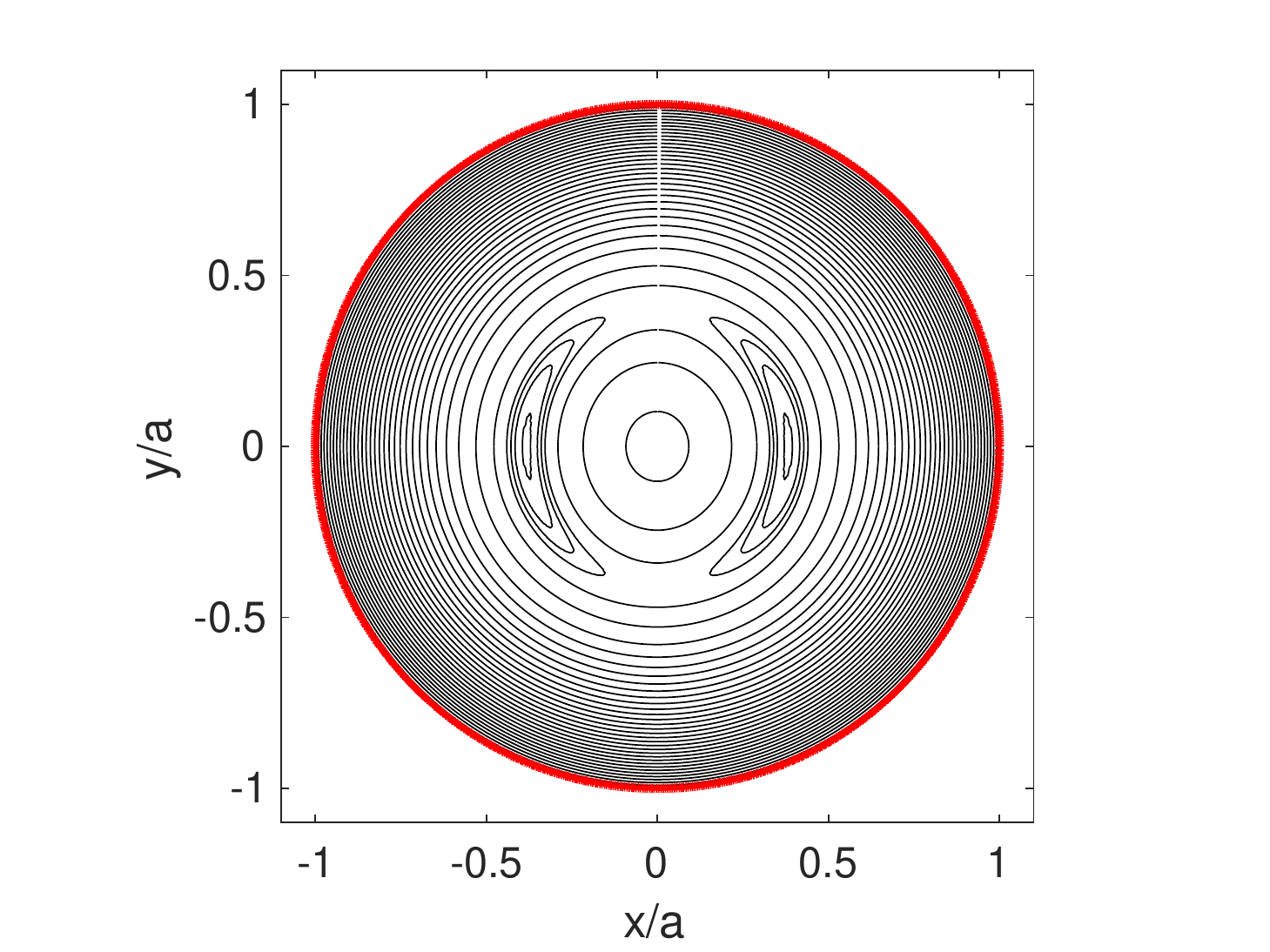} 
\includegraphics[clip=true, scale=0.52]{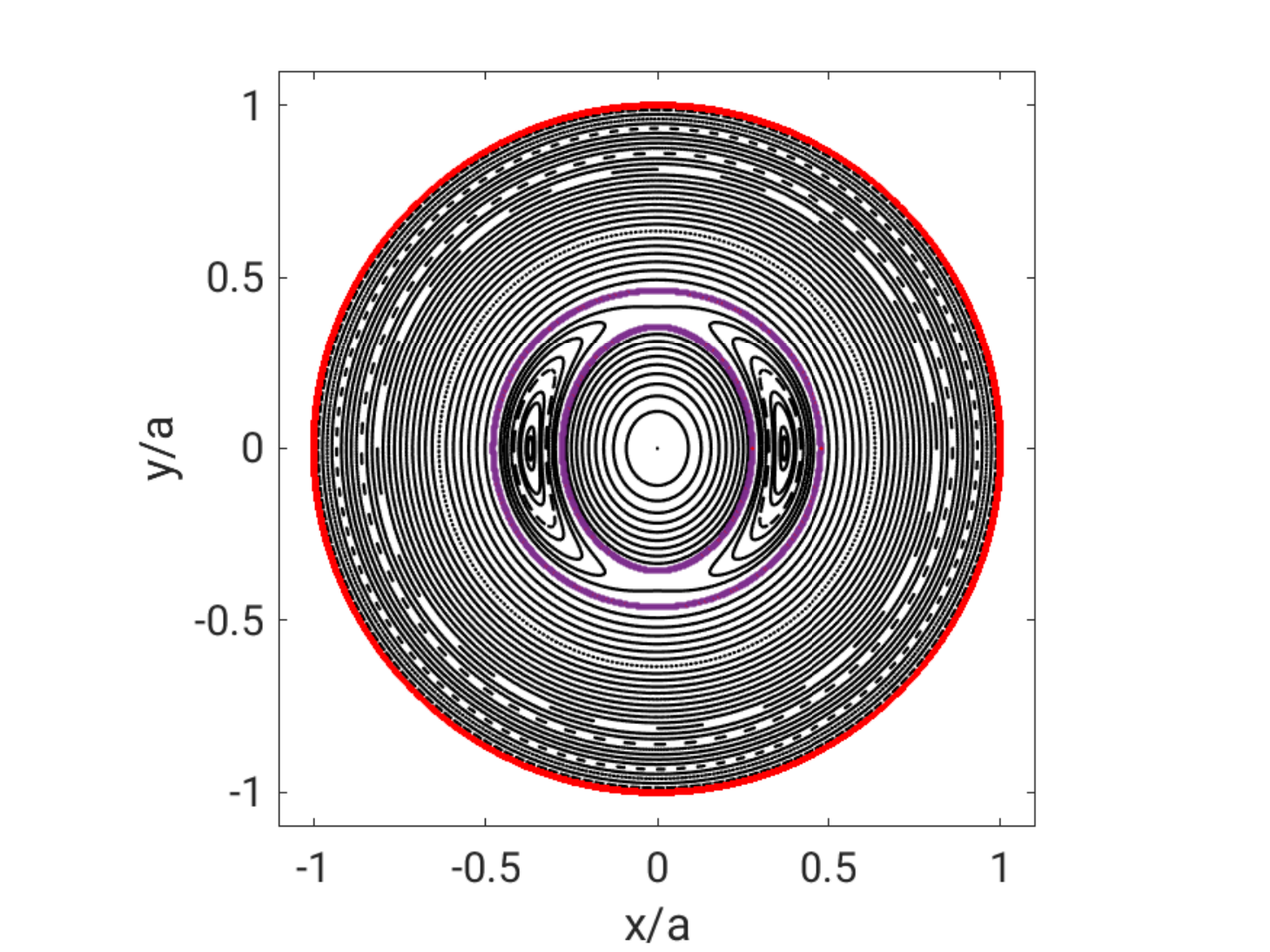}
\caption{Poincaré sections of the magnetic field at $\varphi=0$ showing the saturated island as obtained from SpeCyl (left column) and SPEC (right column) for $q_0=1.3$ (top) and $q_0=1.6$ (bottom). The violet lines on the SPEC Poincaré sections (right column) indicate the two interfaces encapsulating the resonant volume.} \label{fig:poinc} 
\end{figure}

\begin{figure}[h!]
\centering
\includegraphics[clip=true, scale=0.65]{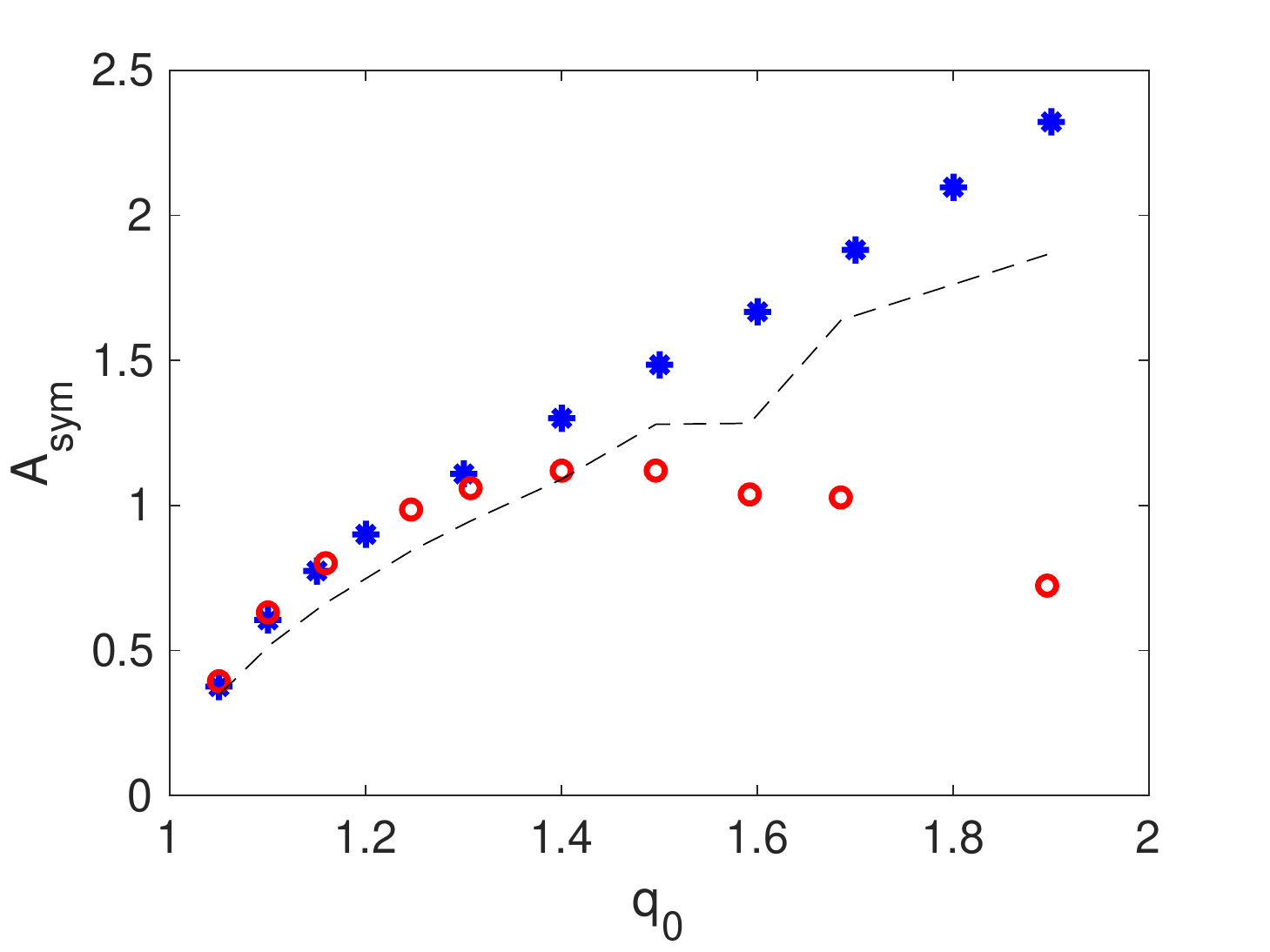}
\caption{$A_{sym}$ as a function of $q_0$ as obtained from SpeCyl (blue stars) and SPEC (red circles). The dashed line indicates the expected island asymmetry $A_{max}$ obtained by solving by solving Eq.~(\ref{cons2}) and that is used to initialize the SPEC equilibrium.} \label{fig:asymspecyl} 
\end{figure}

\begin{figure}[h!]
\centering
\includegraphics[clip=true, scale=0.65]{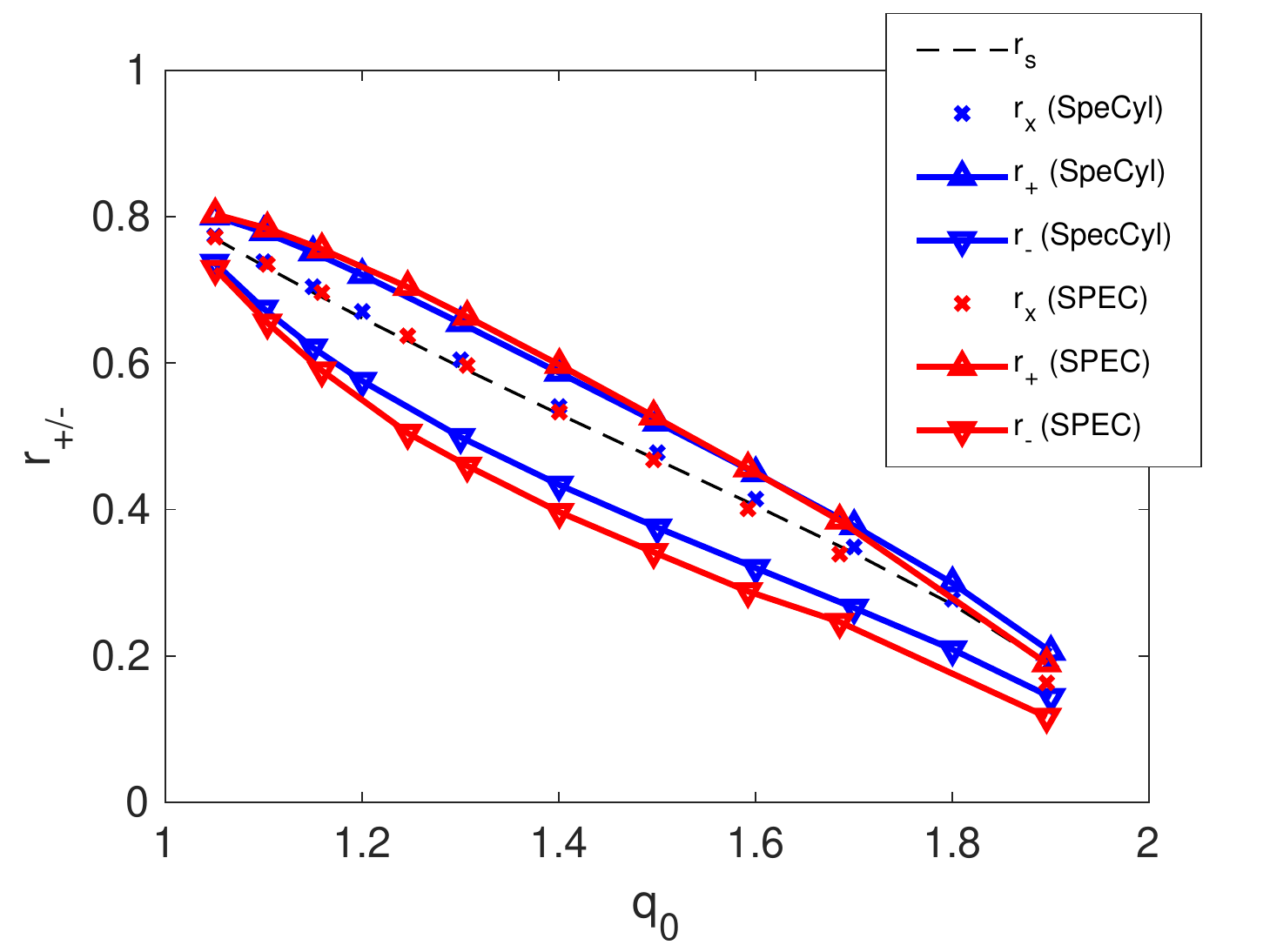}
\caption{Radial positions $r_X$, $r_-$, and $r_+$ characterizing the island, as a function of $q_0$ as obtained from SpeCyl (blue) and SPEC (red). The dashed line indicates the position of the resonant surface.} \label{fig:rpm} 
\end{figure}

We run SpeCyl for a number of initially unstable equilibria that, as described in Sec.~\ref{sec:eq}, are characterized by $q_0$. For each value of $q_0$, we consider different values for the Lundquist number from $S=10^6$ to $S=10^9$ at fixed magnetic Prandtl number $P=S/M=1$, we evolve the system until saturation and measure the island width. Figure \ref{fig:poinc} illustrates examples of such saturated states by showing a Poincar\'e section of the magnetic field obtained for $q_0=1.3$ (top left) and $q_0=1.6$ (bottom left) with $S=10^8$. For sufficiently high Lundquist number $S \geq 10^8$, the radial profiles of the perturbed flux function $\hat{\psi}$ show good agreement in the linear growth phase with those produced by the shooting code, and the nonlinear saturated island width becomes independent of $S$ (data not shown). This is consistent with previous analytical and numerical studies (see e.g. Ref \cite{Poye2014}) showing that the saturated tearing mode island width is independent of resistivity and viscosity. We also find very good agreement with the analytical predictions from the nonlinear theory of Arcis for all the considered tearing unstable equilibria (Figure \ref{fig:wsat}). In particular, the best agreement with nonlinear theory is found assuming $\sigma=1$ (i.e., a uniform equilibrium electric field) consistently with the assumption of the resistive MHD code. From the resistive simulations, we also observe that the saturated island is not symmetric with respect to the X-point. More precisely, we can define the asymmetry parameter  
\begin{equation}
\hspace*{3cm}
A_{sym} = \frac{r_X-r_-}{r_+-r_X}  - 1 \label{asym}
\end{equation}
where $r_X$ is the radial position of the X-point of the island and $r_-$, $r_+$ are respectively the smallest and largest radial positions of the island separatrix (which occur at the poloidal position of the O-point). For a symmetric island, $r_X-r_- = r_+ - r_X$ and thus $A_{sym}=0$. We can evaluate Eq.~(\ref{asym}) from the SpeCyl simulation results and display $A_{sym}$ as a function of $q_0$ (Figure \ref{fig:asymspecyl}). Figure \ref{fig:rpm} also shows how the individual values of $r_X$, $r_-$, and $r_+$ change with $q_0$. As soon as $q_0>1$, the islands become quickly asymmetrical, $A_{sym}>0$, and become more and more asymmetric as $q_0$ approaches 2. We remark that the asymmetric feature of tearing mode islands has been observed in previous investigations and has been suggested to play an important role in determining the saturation amplitude from a modified Rutherford equation approach \cite{Teng2016}.

\begin{figure}[h!]
\centering
\includegraphics[clip=true, scale=0.65]{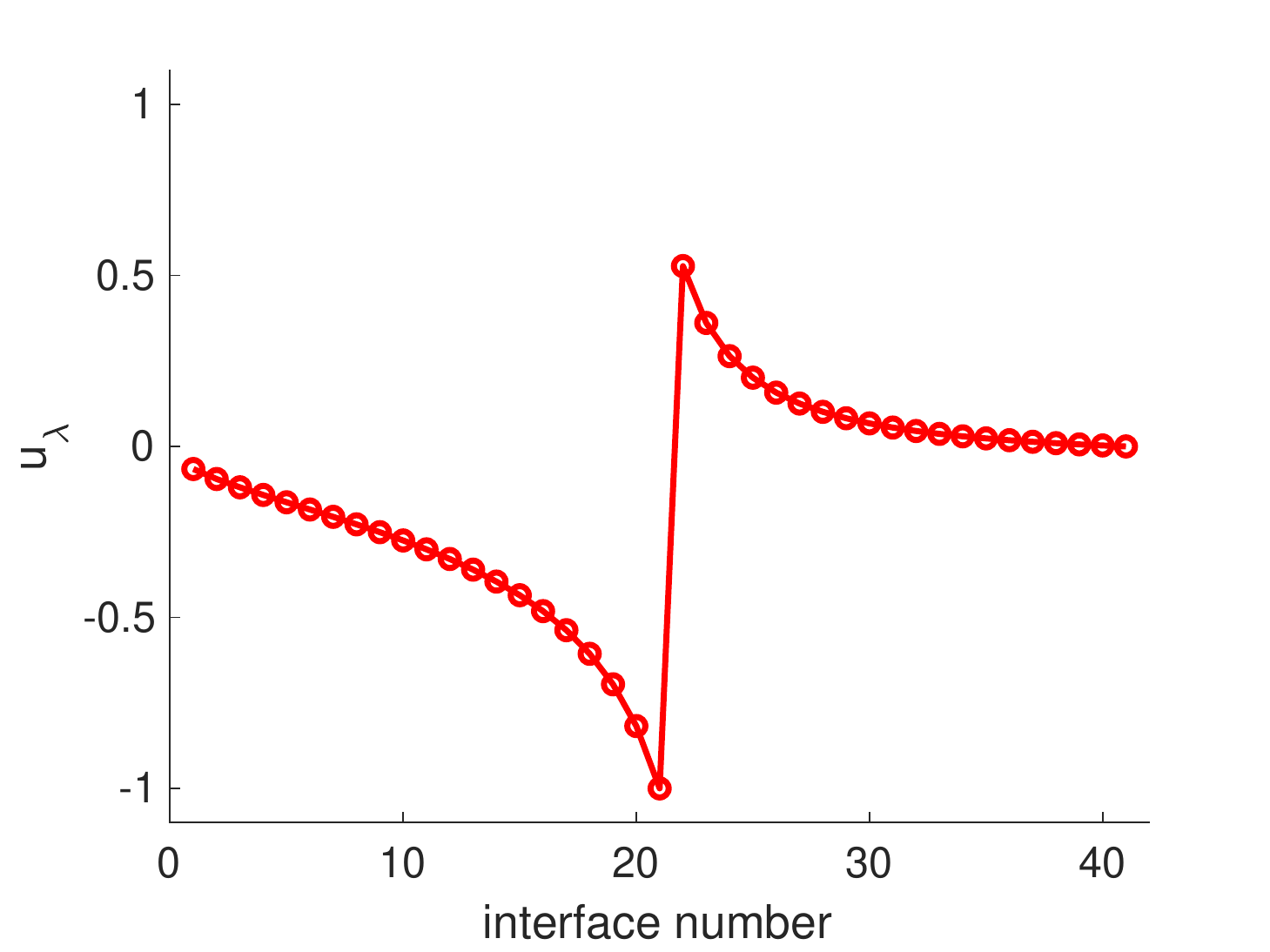}
\caption{Unstable eigenmode profile showing the typical tearing parity around the resonant surface. Here $q_0=1.1$ and $N_v=41$.} \label{fig:xi} 
\end{figure}

\begin{figure}[h!]
\centering
\includegraphics[clip=true, scale=0.65]{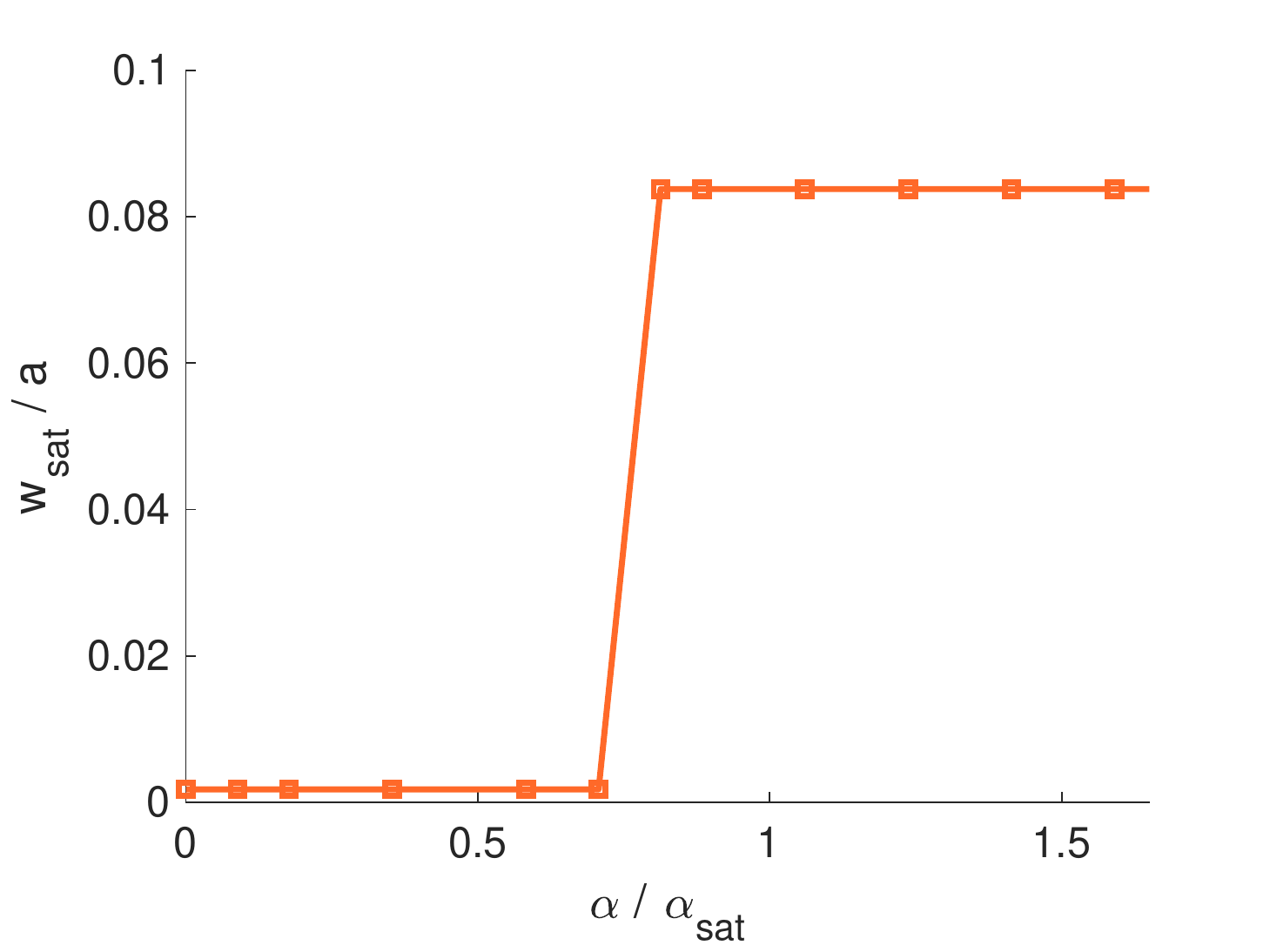}
\caption{Saturated island width obtained from SPEC nonlinear calculations with $q_0=1.1$ as a function of the initial perturbation amplitude as measured by $\alpha$. The normalizing factor $\alpha_{sat}$ represents the actual saturation amplitude of the helical state obtained after the nonlinear calculation. The Fourier resolution is $(m,n)=(4,2)$.} \label{fig:alpha} 
\end{figure}

\section{Nonlinear equilibrium calculations} \label{sec:nlspec}

\begin{figure}[h!]
\centering
\includegraphics[trim={2cm 6cm 2cm 2cm},clip, scale=0.55]{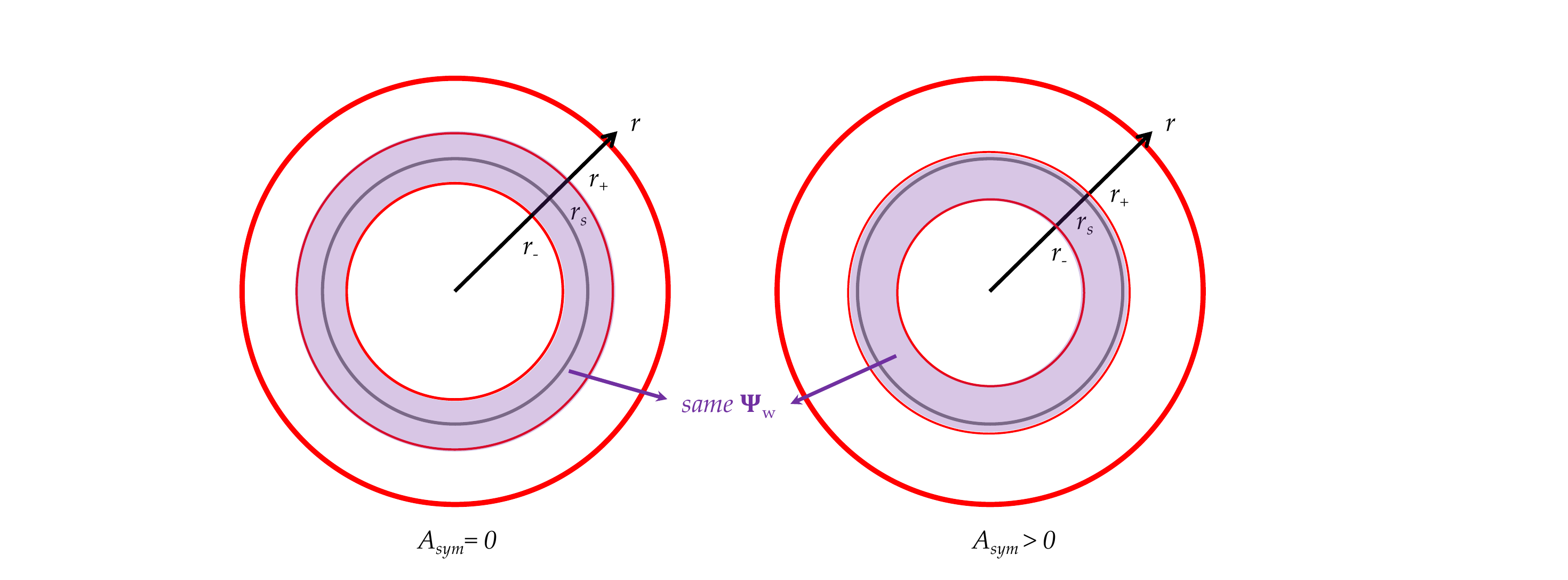}
\caption{An illustration of how the initial positions $r_+$ and $r_-$ of the interfaces defining the resonant volume can be symmetric (left) or asymmetric (right) with respect to the resonant surface $r=r_s$, while preserving the same enclosed toroidal flux $\Psi_w$.} \label{fig:asym} 
\end{figure}

We now use the SPEC code to seek stable equilibrium solutions by starting from the unstable equilibria described in Sec.~\ref{sec:eq}. Indeed, in Sec.~\ref{sec:specstab} we obtained, for each value of $q_0$, the unstable eigenmode $\bv{u}_{\lambda}\equiv\{r_{l,mn}\}$ that is associated with the unstable (negative) eigenvalue $\lambda$ of the force-gradient matrix $\mathcal{G}$. Here $m=2$, $n=1$, and the index $l=1,2,...,N_v-1$ labels interfaces. Figure \ref{fig:xi} shows an example of such an eigeinmode. We now perturb the Fourier geometry of the initial equilibrium interfaces by adding a displacement proportional to the eigenmode, namely, we add $\alpha\ \bv{u}_{\lambda} $ with $\alpha>0$. Notice that, given a certain volume (or equivalently, magnetic flux) enclosed by the two interfaces surrounding the resonant surface, there is a maximum perturbation amplitude beyond which the interfaces would overlap ––  but this is never exceeded because the perturbation is typically small enough. SPEC then seeks a new equilibrium solution. If $\alpha$ is too small, SPEC snaps back to the initial equilibrium (Figure \ref{fig:alpha}). If $\alpha$ is large enough, however, SPEC finds another equilibrium solution (independent of $\alpha$, see Fig.~\ref{fig:alpha}) that has a helical structure and displays a magnetic island inside the volume containing the resonant surface and thus where reconnection happens (two examples are shown in Fig.~\ref{fig:poinc}). This approach was successful in slab geometry and showed that the saturation amplitude of the tearing mode is correctly retrieved \cite{Loizu2020}. In that study, the toroidal magnetic flux $\Psi_w$ enclosed by the resonant volume, which constrains the maximum achievable island width, was chosen by following a recipe without free parameters that is summarized in Sec.~\ref{sec:psiw} for convenience. If the island is expected to be perfectly symmetric around the resonant surface, the value of $\Psi_w$ uniquely determines the initial positions $r_+$ and $r_-$ of the two interfaces defining the resonant volume, since $\Psi_w = B_z \pi(r_+^2 - r_-^2)$ and the symmetry argument forces the flux to be equally distributed on each side of the resonant radius $r_s$. In general, however, there are multiple ways of placing the two interfaces given the same total flux $\Psi_w$ (see Fig.~\ref{fig:asym} for an illustration). For this reason, we need an additional constraint that is related to the expected island asymmetry, and that is the subject of Sec.~\ref{sec:asym}. Finally, the search for a nearby, lower energy equilibrium state can be done with different constraints, and an appropriate choice must be made by identifying the good quasi-invariants in the process at play. In the case of tearing mode reconnection, it is the net-toroidal-current profile that shall be constrained, and that is the subject of Sec.~\ref{sec:jtor}. Section \ref{sec:wspec} summarizes the results of nonlinear simulations carried out with SPEC to reproduce the saturated tearing mode island width as a function of $q_0$.

\subsection{Determining $\Psi_w$} \label{sec:psiw}

\begin{figure}[h!]
\centering
\includegraphics[clip=true, scale=0.75]{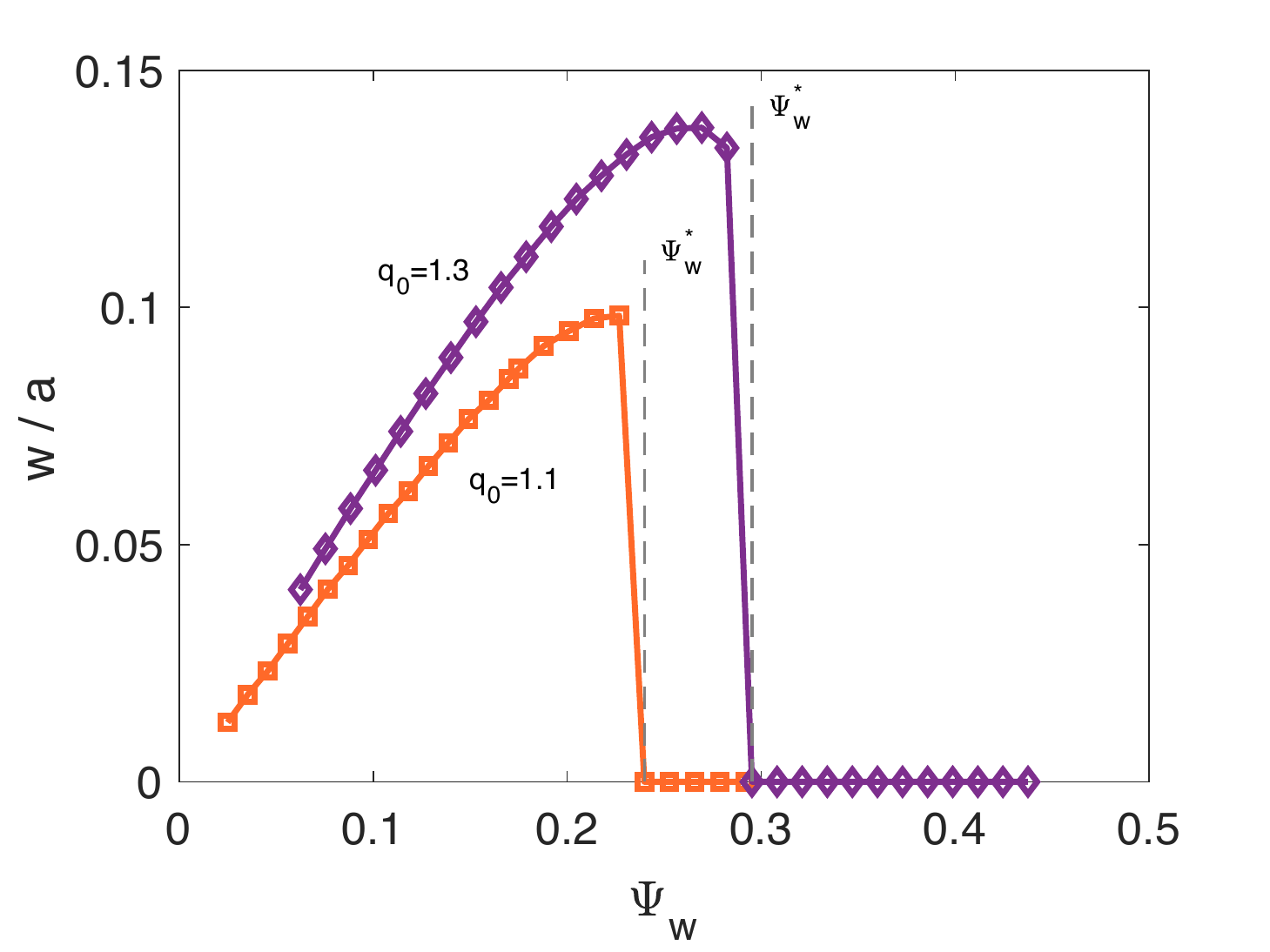}
\caption{Island width, $w$, obtained from SPEC nonlinear calculations as a function of the enclosed toroidal flux $\Psi_w$ imposed in the resonant volume. Two values of $q_0$ are shown: $q_0=1.1$ (orange squares) and $q_0=1.3$ (purple diamonds). The vertical dashed lines indicate the value of $\Psi_w^*$. Here, the flux is always equally distributed on each side of the resonant surface, and the toroidal current profile is constrained. The Fourier resolution is low here ($m=2,n=1$).} \label{fig:wvp} 
\end{figure}

\begin{figure}[h!]
\centering
\includegraphics[clip=true, scale=0.7]{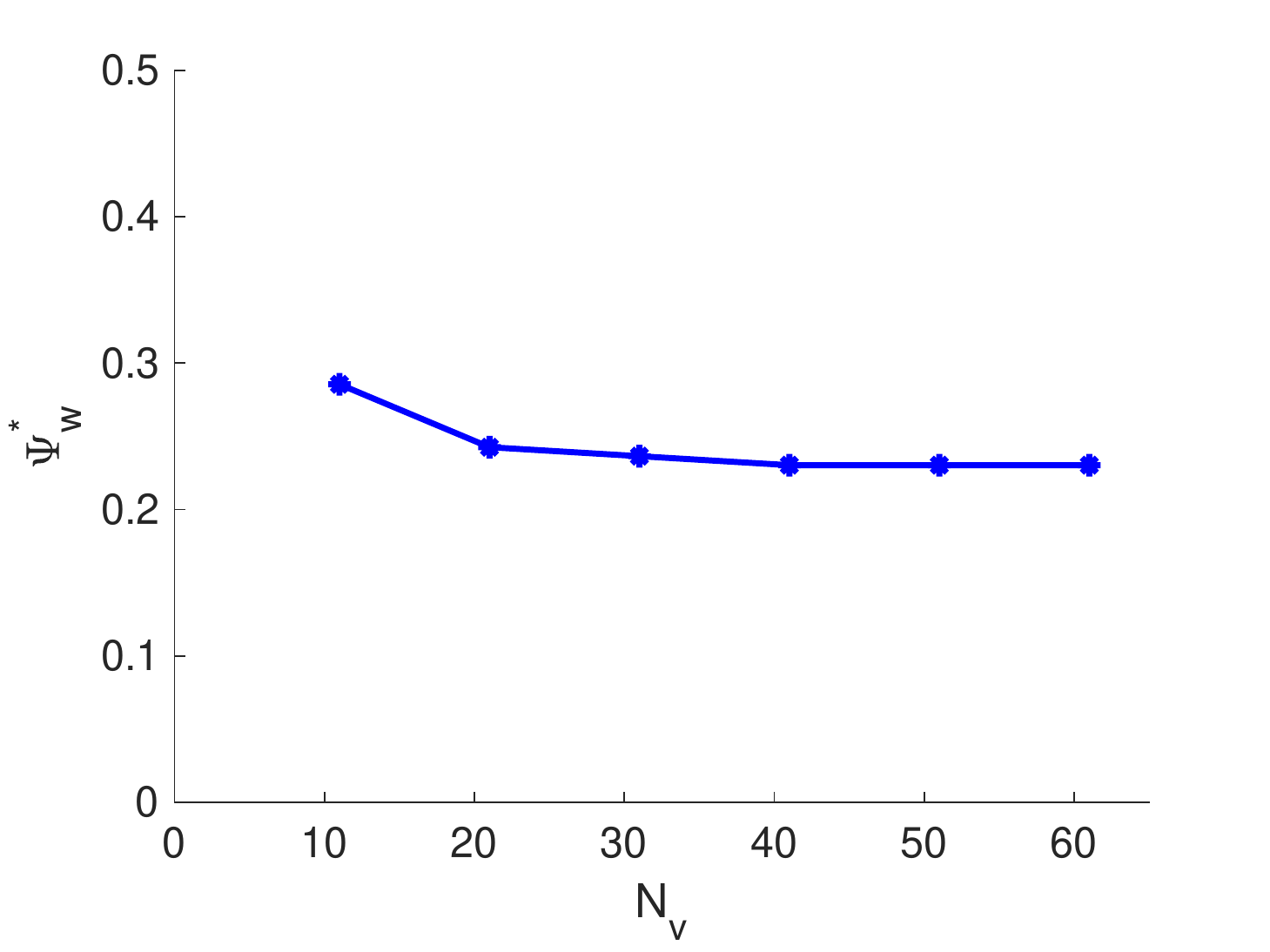}
\caption{Values of $\Psi_w^*$ ( obtained by solving $\lambda(\Psi_w)=0$) as a function of the number of volumes. Here the equilibrium has $q_0=1.1$.} \label{fig:nvolconv} 
\end{figure}

In the limit $\Psi_w\rightarrow0$, the volume available for reconnection vanishes and therefore the maximum island width that is possibly achievable goes to zero. For increasingly large values of $\Psi_w$, the island is allowed to grow over a larger volume, although it may well only occupy a fraction of it. However, if the value of $\Psi_w$ exceeds a certain threshold, $\Psi_w^*$, the initial equilibrium becomes stable and no island grows at all. This is illustrated in Figure \ref{fig:wvp} for two cases. Obviously, for each $q_0$, the saturated island width obtained with SPEC depends on $\Psi_w$. However, there is a distinct point, which we identify as $w_{sat}$, and that is the maximum achieved island width, which occurs at around $\Psi_w\approx\Psi_w^*$. We thus apply the following procedure: for each value of $q_0$, a linear stability analysis is first performed with SPEC (see Sec.~\ref{sec:specstab}) and the value of $\Psi_w^*$ is obtained by solving $\lambda(\Psi_w)=0$. This criterion is reminiscent of the quasi-linear theory of Roscoe White \cite{White1977}, in which the saturation of a tearing mode occurs at a marginal stability point evaluated as $\Delta^\prime(w)=0$. We remark that the value of $\Psi_w^*$ obtained with this procedure is independent of the number of volumes for sufficiently large $N_v$ (see Fig. \ref{fig:nvolconv}). Then, SPEC is run nonlinearly with $\Psi_w \lesssim \Psi_w^*$ and the obtained value of $w_{sat}$ is extracted from the corresponding saturated state. 

\subsection{Model for the island asymmetry} \label{sec:asym}

The SPEC calculations shown in Fig.~\ref{fig:wvp} were done by assuming that the flux $\Psi_w$ in the resonant volume is equally distributed on each side of the resonant surface. However, a different choice (as illustrated in Fig.~\ref{fig:asym}) can substantially modify the nonlinear saturation amplitude and thus the value of $w_{sat}$. In the following, we describe how we can estimate the expected asymmetry of the island and thus constrain accordingly the initial position of the interfaces, $r_+$ and $r_-$, around the resonant surface.

We start by writing the total magnetic field as $\bv{B} = \bv{B}_0(r) + \nabla\psi\times\hat{\bv{z}}$ and assume an arbitrarily large perturbation with single helicity, $\psi(r,\zeta)=\hat{\psi}(r)\cos{\zeta}$, where $\zeta=m\theta-n\varphi$. This perturbation generates an island around the rational surface with $q_s = q(r_s)=m/n$. The X-point of the island is located at $\zeta=0$ and the O-point of the island is located at $\zeta=\pi$. However, as we show now, this island is not symmetric in the sense of Eq.~(\ref{asym}), namely $A_{sym}>0$. The island itself consists of a set of nested magnetic surfaces that are described by a flux-function $\chi(r,\zeta)$ such that $\bv{B}\cdot\nabla\chi=0$. One can show that \cite{Fitzpatrick1995}
\begin{equation}
\chi(r,\zeta)=\int_r^{r_s}\left( 1-\frac{q}{q_s} \right)B_{\theta}dr + \hat{\psi}(r)\cos{\zeta} \label{chi} 
\end{equation}
and so the magnetic surfaces in the island, and in particular the island separatrix, are surfaces of constant $\chi$. Furthermore, the asymptotic behaviour of $\hat{\psi}(r)$ around the resonant surface can be obtained from the linear ideal outer solutions (see Section 6.9 of Ref.~\cite{Wesson2004}) and is given by 
\begin{equation}
\hat{\psi}^\pm(r) =  \psi_s \left[ 1 + \mathcal{A}(r-r_s)\ln{|r-r_s|} + A^{\pm}(r-r_s))  \right] \label{hatpsi} 
\end{equation}
where $\pm$ refers to each side of the resonant surface, $\psi_s$ is the (unknown) amplitude of the perturbation at the resonant surface, $\mathcal{A}$ is given by Eq.~(\ref{arcispar}), and $A^{\pm}$ is a constant that can be determined numerically by solving for the radial profile of $\hat{\psi}(r)$ outside of the resonant layer. As a matter of fact, the constants $A^{\pm}$ are used to evaluate $\Delta^\prime$ and $\Sigma'$ as defined in Eqs.~(\ref{dp}) and (\ref{sp}), since $\Delta^\prime=A^+ - A^-$ and $\Sigma'=A^+ + A^-$. At the separatrix, we have
\begin{equation}
\chi(r_+,\pi)=\chi(r_-,\pi)\equiv \chi_{sep} \approx \chi(r_s,0) = \psi_s  \label{chisep}
\end{equation}
and hence combining Eqs.~(\ref{chi}) and (\ref{hatpsi}) into Eq.~(\ref{chisep}) we obtain a constraint for $(r_+, r_-)$,
\begin{equation}
g(r_+) = g(r_-) \label{cons2} 
\end{equation}
where 
\begin{equation}
g(r) = \frac{\int_r^{r_s}\left( 1-\frac{q}{q_s} \right)B_{\theta}dr }{2+\mathcal{A}(r-r_s)\ln{|r-r_s|} + A^{\pm}(r-r_s)}
\end{equation}
is a known function of the equilibrium. Equation (\ref{cons2}) can be combined with the constraint for the toroidal flux, $\Psi_w=\pi B_{z0}(r_+^2-r_-^2)$, to solve for the positions $r_+$ and $r_-$ of the interfaces that encapsulate the resonant volume. The values obtained for $r_+$ and $r_-$ can then be used to initialize the SPEC equilibrium interfaces. Given this initial geometry, one can show that the expected (maximum) value of the island asymmetry at saturation is
\begin{equation}
A_{max} = 2 \left(\frac{r_s-r_-}{r_+-r_s}-1\right)   \label{Amax}   
\end{equation}
The value of $A_{max}$ obtained by solving Eq.~(\ref{cons2}) for different values of $q_0$ is shown in Fig.~\ref{fig:asymspecyl} and the agreement with the actual saturated island asymmetry obtained by SpeCyl is quite remarkable. Notice that this prediction is merely based on linear theory.

\subsection{Constraining the current profile} \label{sec:jtor}

Magnetic reconnection is a process through which magnetic energy is converted into kinetic and thermal energy. If the process is fast enough, the magnetic helicity is well conserved \cite{Berger1999}, and this happens for example during sawteeth in tokamaks \cite{Heidbrink2000}, which occur on quasi-ideal time scales. Resistive tearing modes, however, are instabilities that promote magnetic reconnection at a rate that is fast when compared to the resistive time scale $\tau_{\eta}$, but slow when compared to the Alfvenic time scale $\tau_A$. The reconnection process is then too slow for the magnetic helicity to be a good invariant. Nevertheless, it can be shown that the net-toroidal-current is a flux function at saturation and is given by \cite{Militello2004, Escande2004}

\begin{equation}
j_z(\psi) = \langle j_{eq} \rangle_{\psi} \label{jzpsi}
\end{equation}
where $\langle \cdot \rangle_\psi$ is the average over a flux-surface and $\psi(r,\theta,\varphi)$ is the saturated flux-function. Equation (\ref{jzpsi}) implies the quasi-invariance of the current profile, the invariance being exact in the limit of small island width. We can thus exploit this constraint on the current profile and run SPEC nonlinear equilibrium calculations by constraining the current in each volume, $I_{\mathrm{vol}}$, to remain constant, and the surface current on each interface, $I_{\mathrm{surf}}$, to remain zero, thereby ensuring that the current profile remains smooth. 

\subsection{Nonlinear simulation results} \label{sec:wspec}

\begin{figure}[h!]
\includegraphics[clip=true, scale=0.9]{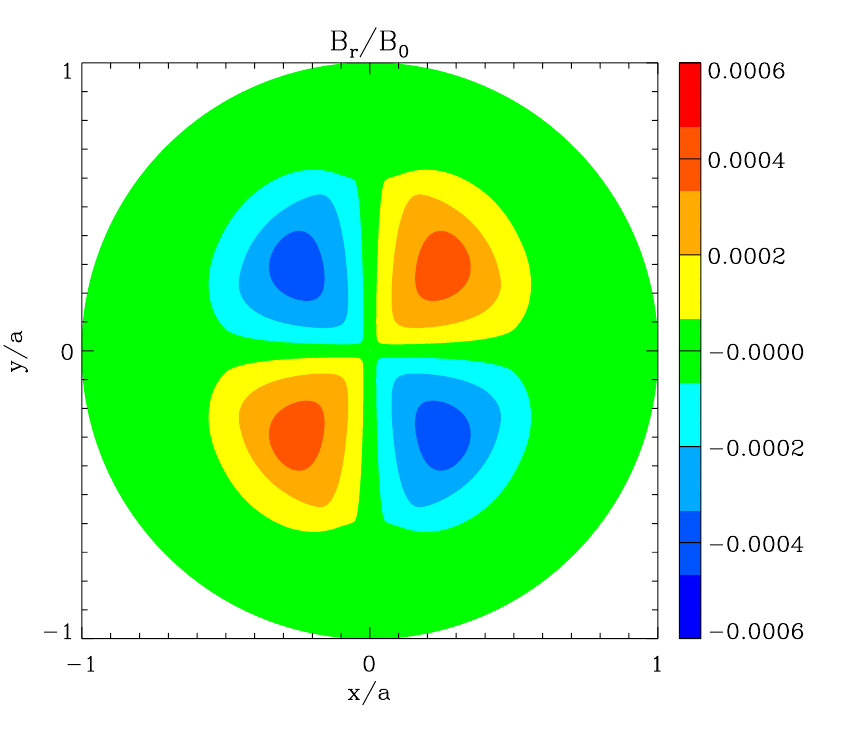} 
\includegraphics[clip=true, scale=0.59]{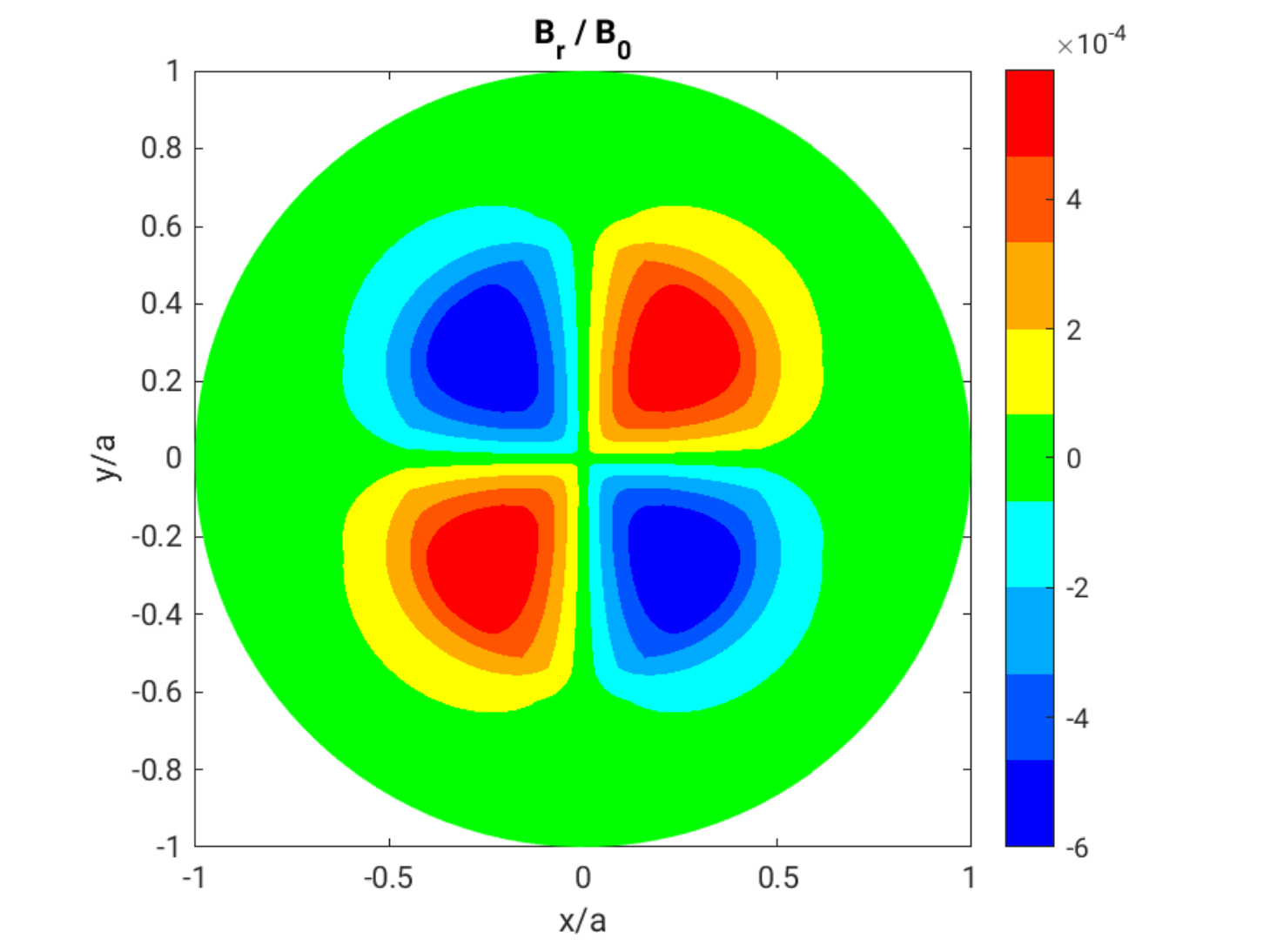} 
\includegraphics[clip=true, scale=0.9]{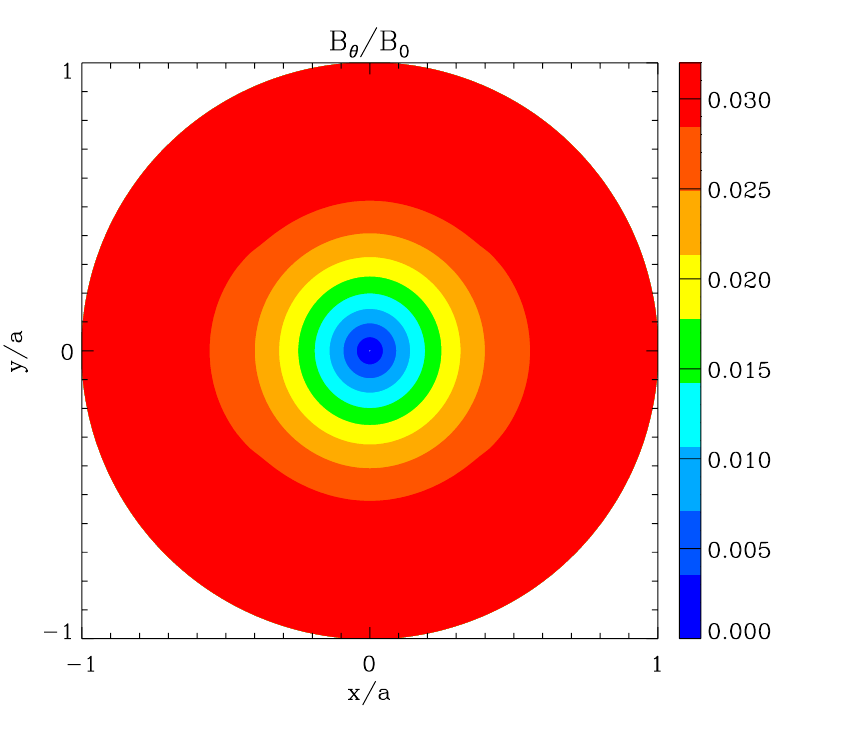}
\includegraphics[clip=true, scale=0.59]{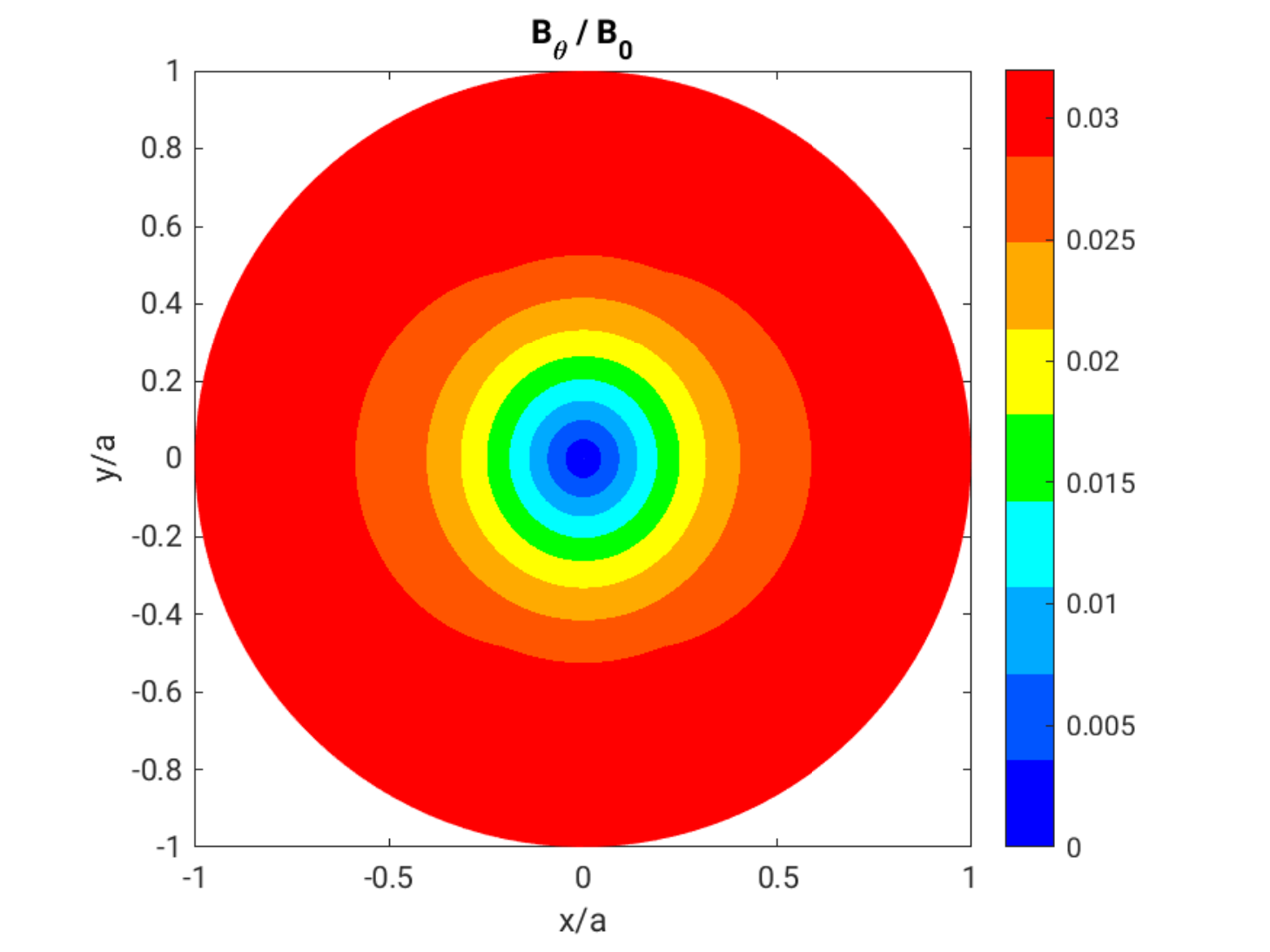}
\includegraphics[clip=true, scale=0.9]{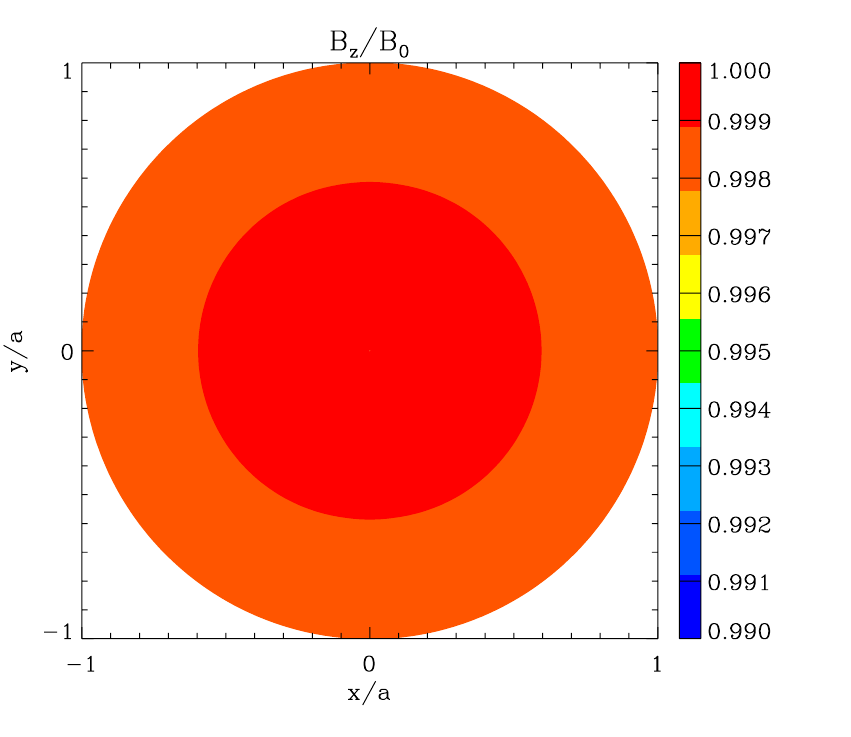}
\includegraphics[clip=true, scale=0.59]{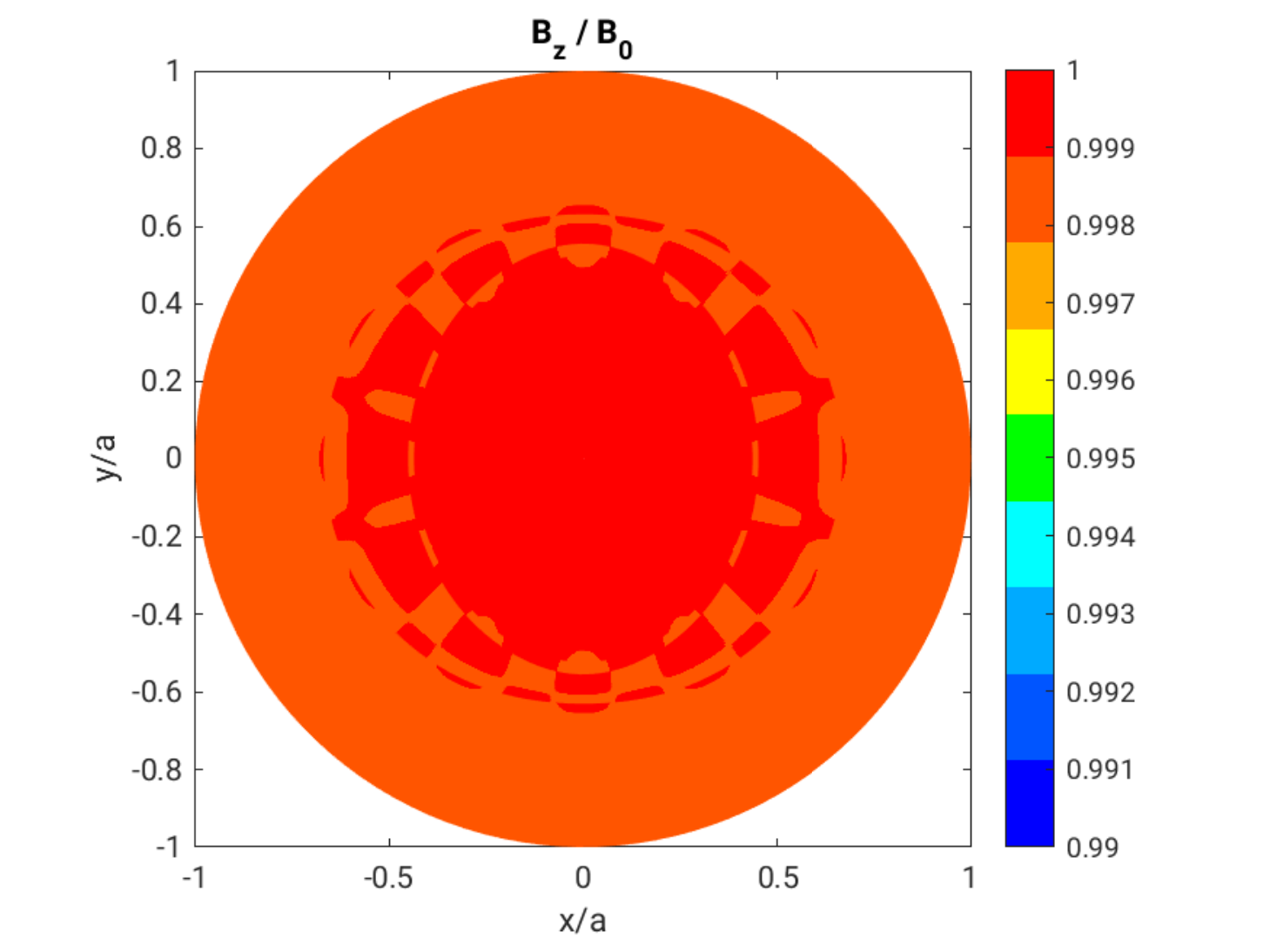}
\caption{Components of $\bv{B}$ at saturation as obtained from SpeCyl (left column) and SPEC (right column) for $q_0=1.3$. For each component, the colorscales are the same between SpeCyl and SPEC.}  \label{fig:Bcomp}
\end{figure}

\begin{figure}[h!]
\includegraphics[clip=true, scale=0.9]{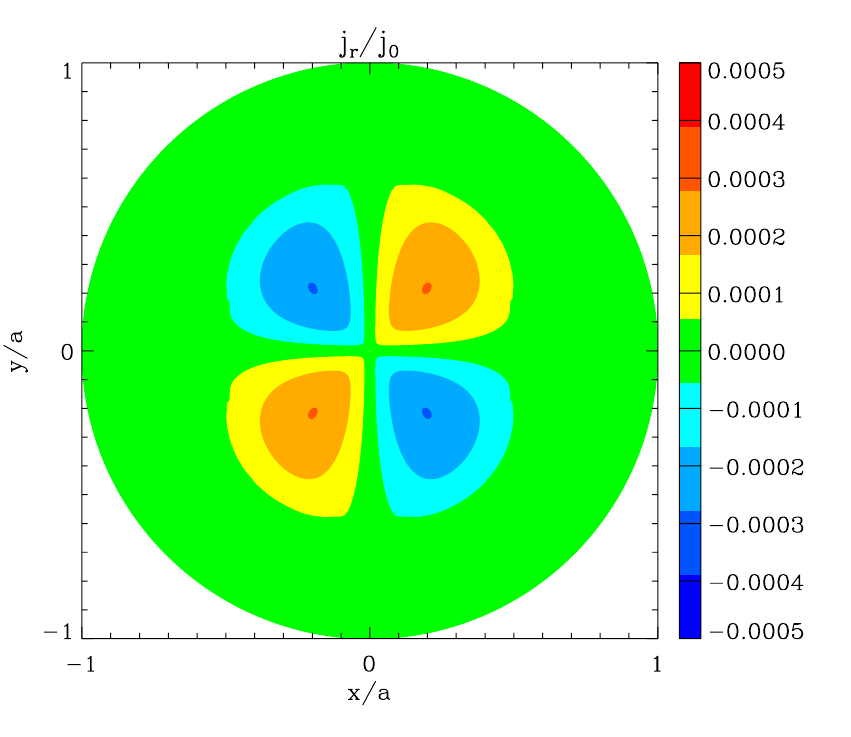} 
\includegraphics[clip=true, scale=0.59]{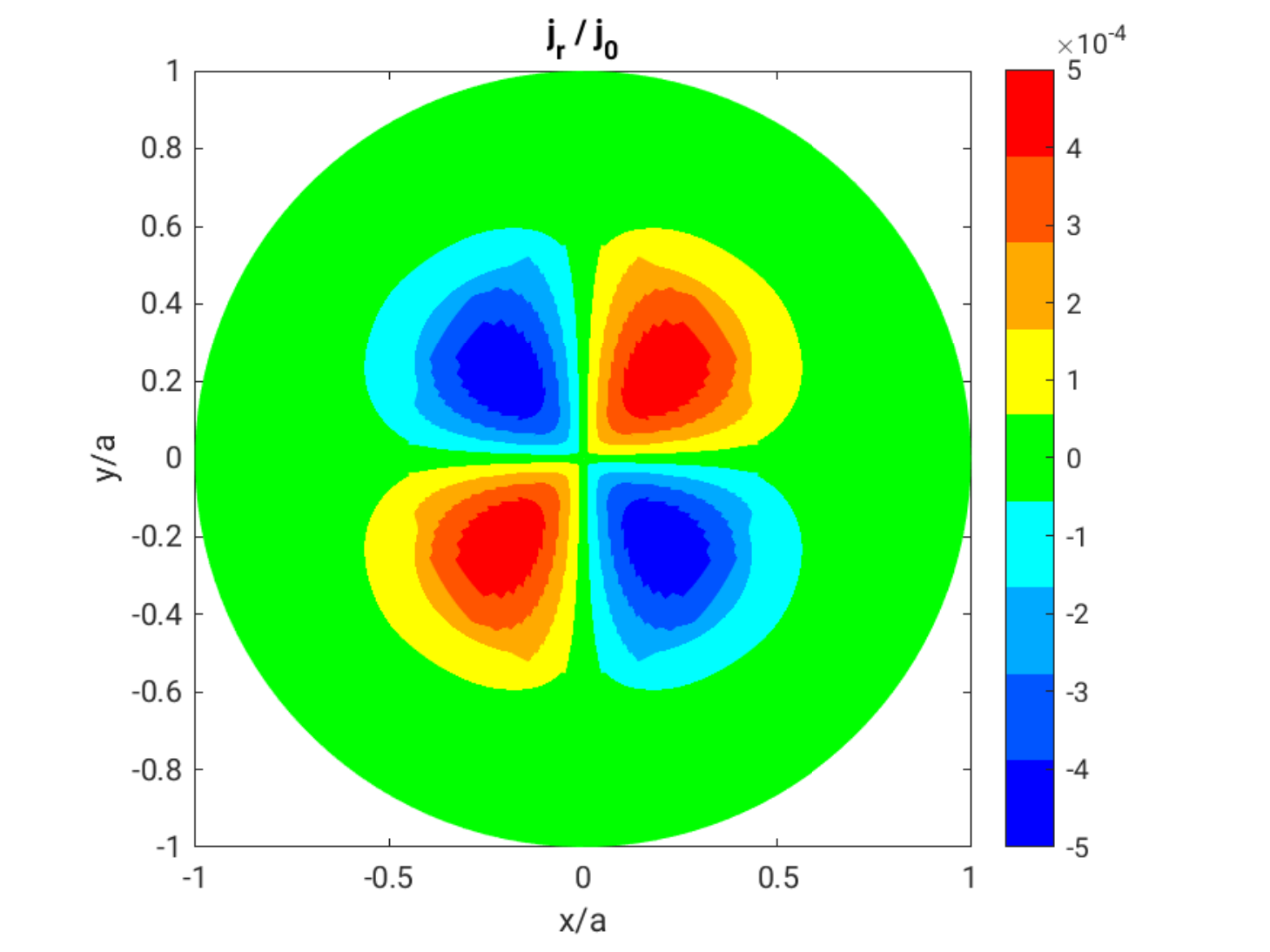} 
\includegraphics[clip=true, scale=0.9]{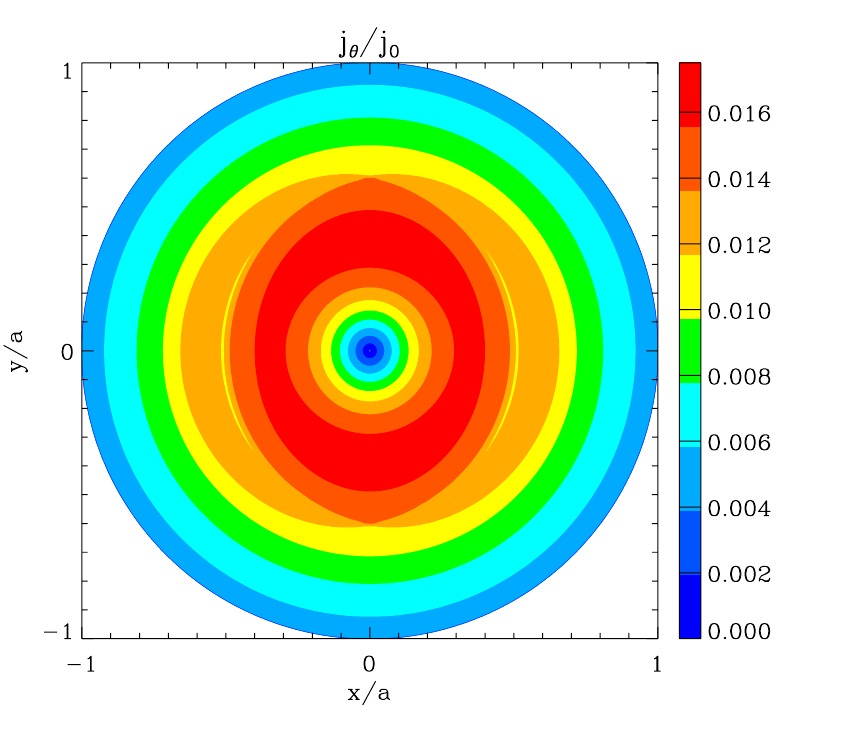}
\includegraphics[clip=true, scale=0.59]{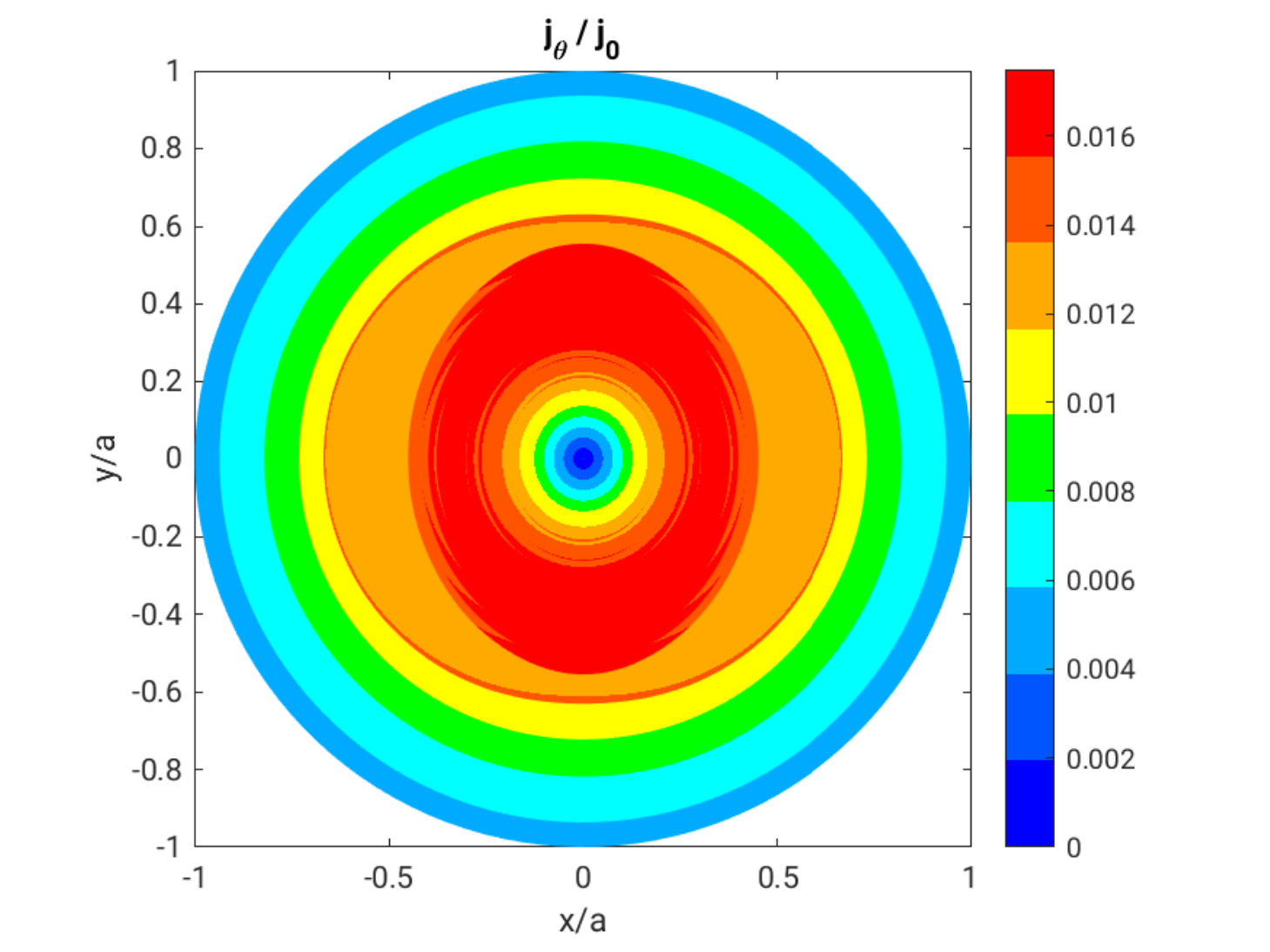}
\includegraphics[clip=true, scale=0.9]{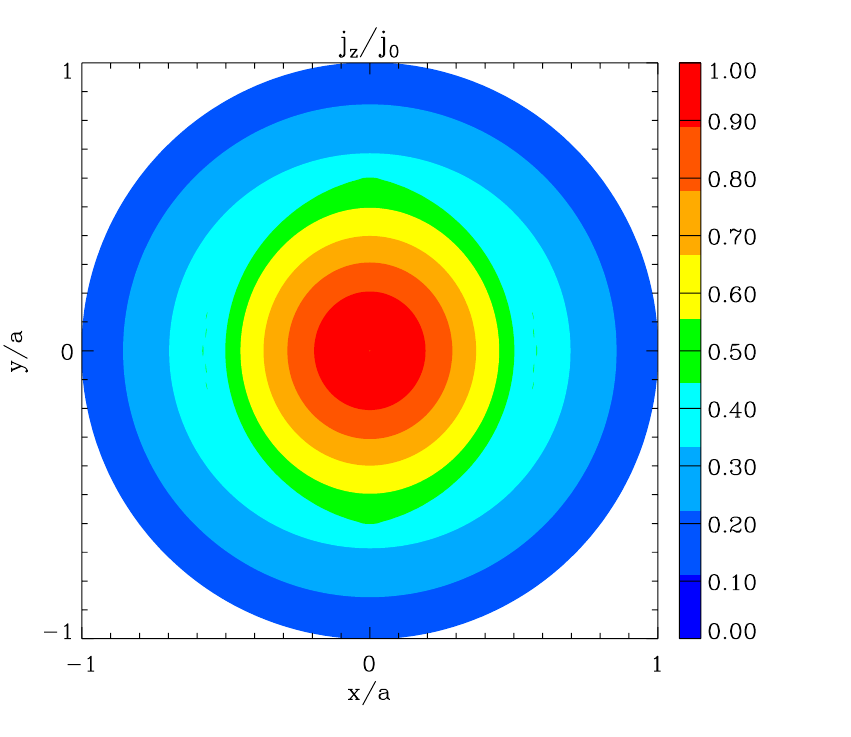}
\includegraphics[clip=true, scale=0.59]{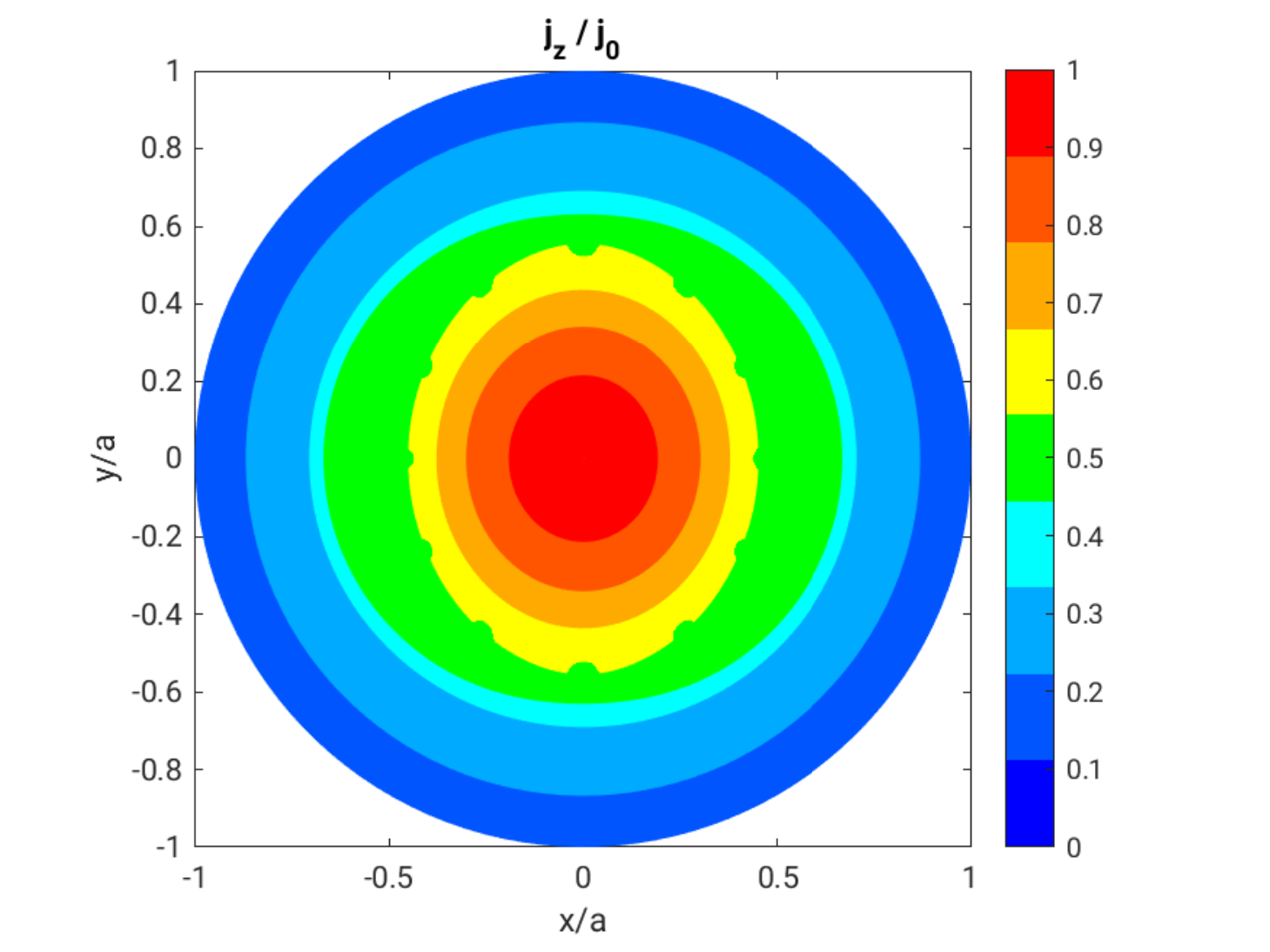}
\caption{Components of $\bv{j}$ at saturation as obtained from SpeCyl (left column) and SPEC (right column) for $q_0=1.3$. For each component, the colorscales are the same between SpeCyl and SPEC.}  \label{fig:jcomp}
\end{figure}

\begin{figure}[h!]
\centering
\includegraphics[clip=true, scale=0.52]{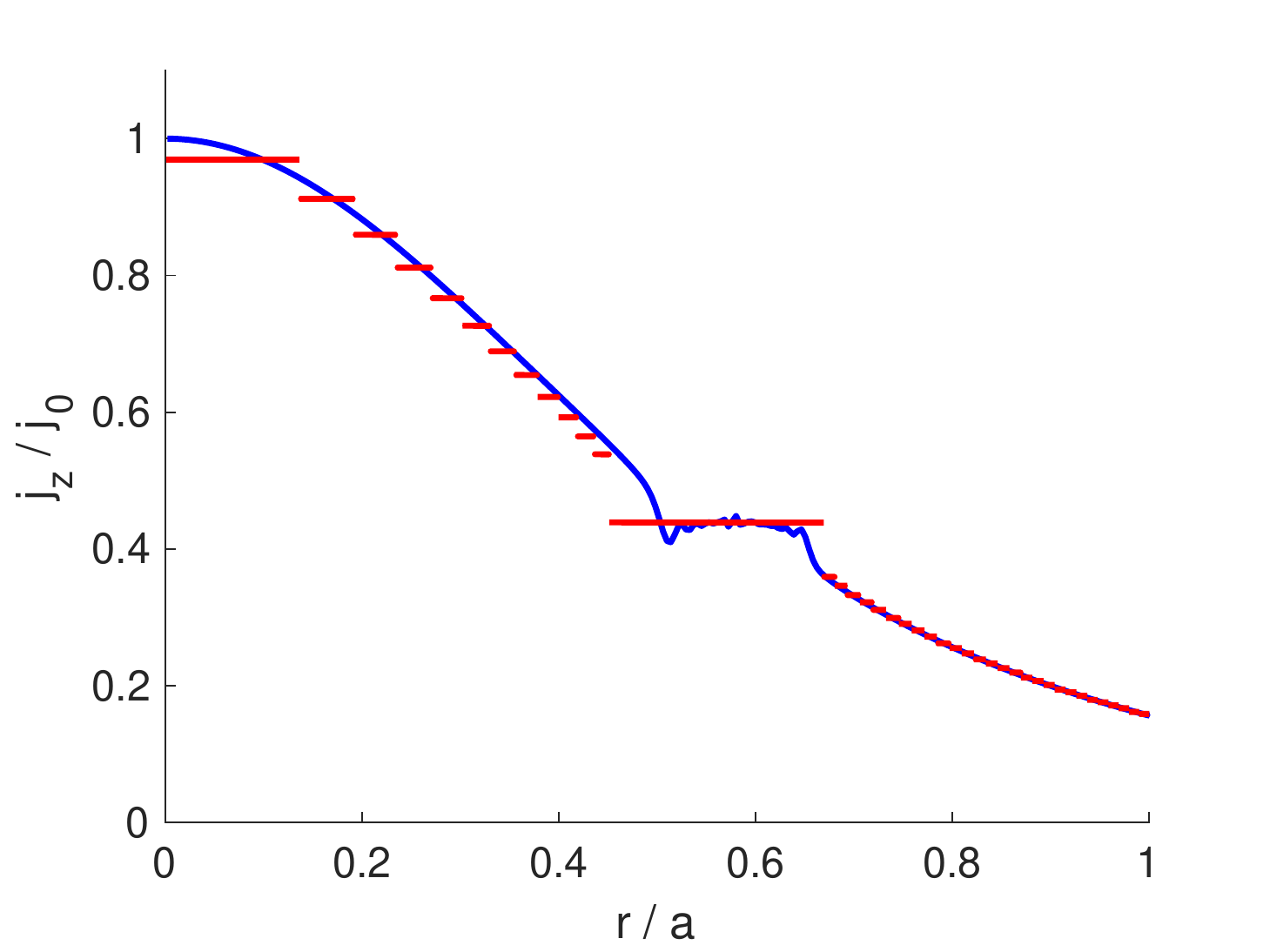} 
\includegraphics[clip=true, scale=0.52]{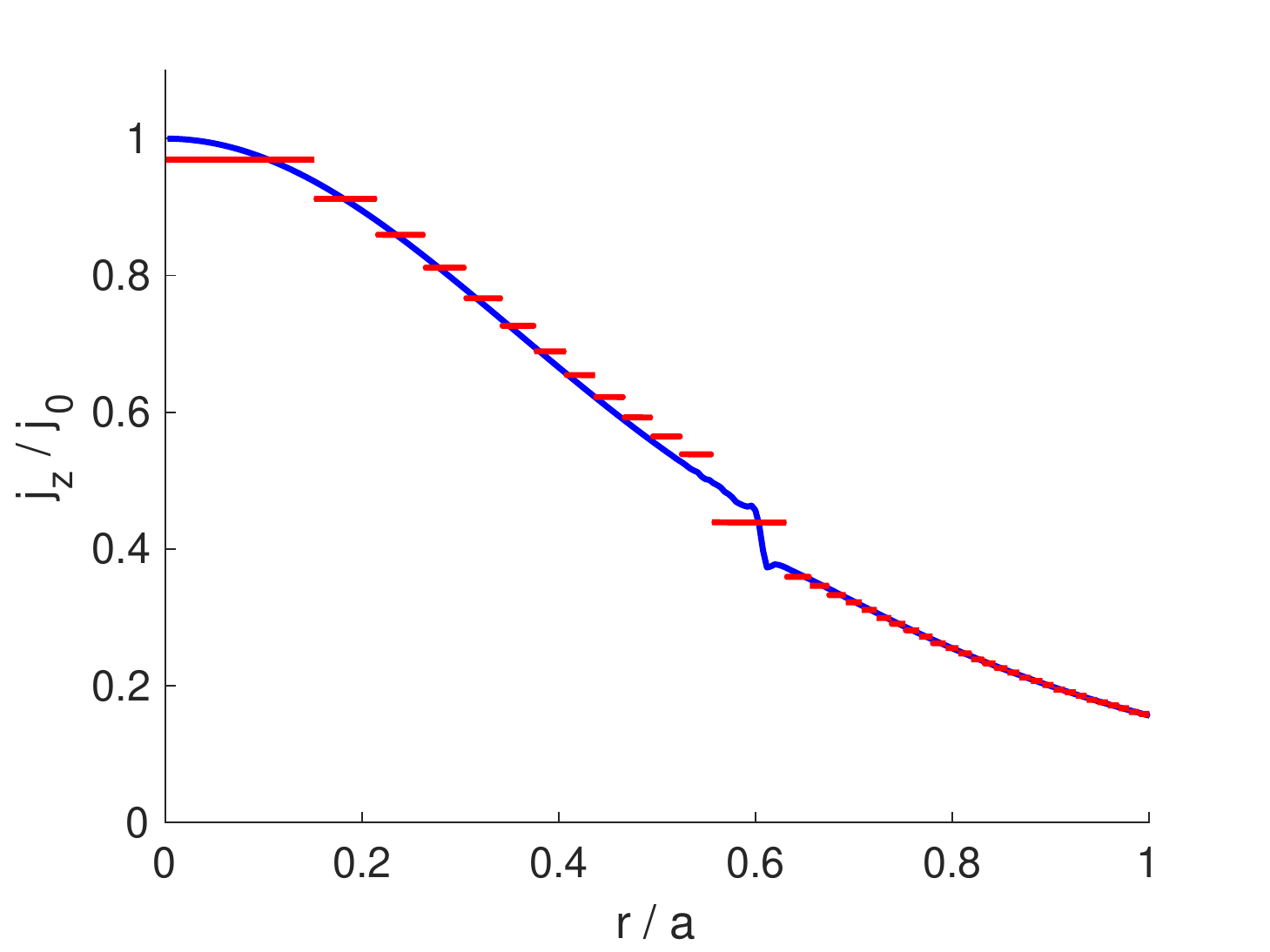} 
\caption{Comparison of the axial current profiles at saturation (red is SPEC, blue is SpeCyl) taken through the O-point (left) and X-point (right) of the island.}  \label{fig:j1d}
\end{figure}

For each value of $q_0$, we thus proceed as follows: (i) perform a linear stability analysis with SPEC to find the unstable eigenmode $\bv{u}_{\lambda}$ and the value of $\Psi_w^*$ (Sec.~\ref{sec:psiw}); (ii) use Eq.~(\ref{cons2}) to infer the values of $(r_+,r_-)$ (Sec.~\ref{sec:asym}); (iii) perturb the equilibrium interfaces according to $\bv{u}_{\lambda}$ and use SPEC to seek a new equilibrium by constraining the current profile (Sec.~\ref{sec:jtor}); (iv) measure the width of the island in the resonant volume of the newly found equilibrium. We remark that this procedure has no free parameters. Figure \ref{fig:poinc} shows two examples of saturated states obtained with SPEC for $q_0=1.3$ (top right) and $q_0=1.6$ (bottom right), which qualitatively look very similar to the ones obtained with SpeCyl (bottom left and top left). Figure \ref{fig:wsat} shows the saturated island width $w_{sat}$ obtained with SPEC as a function of $q_0$. The qualitative shape and order of magnitude of the expected function $w_{sat}(q_0)$ is reproduced. For small enough islands, the agreement with the predictions from nonlinear theory and resistive MHD simulations is remarkable. The agreement remains good up to island sizes of the order of $10\%$ of the minor radius. For increasingly large islands, the equilibrium calculations overestimate the saturation amplitude from nonlinear theory and resistive MHD simulations (the maximum relative error is about $25\%$). This might be due to the fact that these calculations assume a flat current profile inside the island. Furthermore, the initial positioning of the interfaces around the rational surface, which is determined by solving Eq.(\ref{cons2}), has some approximations, namely that the nonlinear perturbed flux function around the resonant surface has a single helicity and a radial structure given by the linear solution. As a matter of fact, we can measure \emph{a posteriori} the positions $r_X$, $r_-$, and $r_+$ that characterize the island at saturation in SPEC, and compare them to those obtained with SpeCyl for different values of $q_0$. This is shown in Fig.~\ref{fig:rpm}. In particular, we see that the sign of the asymmetry is well retrieved, namely the property that $r_X - r_- > r_X - r_+$. However, the exact value of $A_{sym}$, Eq.~(\ref{asym}), is quite sensitive to small errors in the values of $r_X$, $r_-$, and $r_+$, and a good agreement is only observed for $q_0\lesssim 1.3$ (see Fig. \ref{fig:asymspecyl}). We also compare the profiles of each component of $\bv{B}$ and $\bv{j}$ at saturation as obtained from SPEC and SpeCyl. Figures \ref{fig:Bcomp} and \ref{fig:jcomp} show this comparison for the case $q_0=1.3$, for which the agreement in the island width is the least good. Even for this case, we still observe that SPEC is capable of reproducing (both qualitatively and quantitatively) the structure of the saturated tearing mode. We can also compare the axial current profile at saturation, $j_z(r)$, taken through the O-point and X-point of the island (Figure \ref{fig:j1d}). In particular, we indeed see that in SpeCyl the current significantly flattens, but not perfectly, across the island.

\begin{figure}[h!]
\centering
\includegraphics[clip=true, scale=0.65]{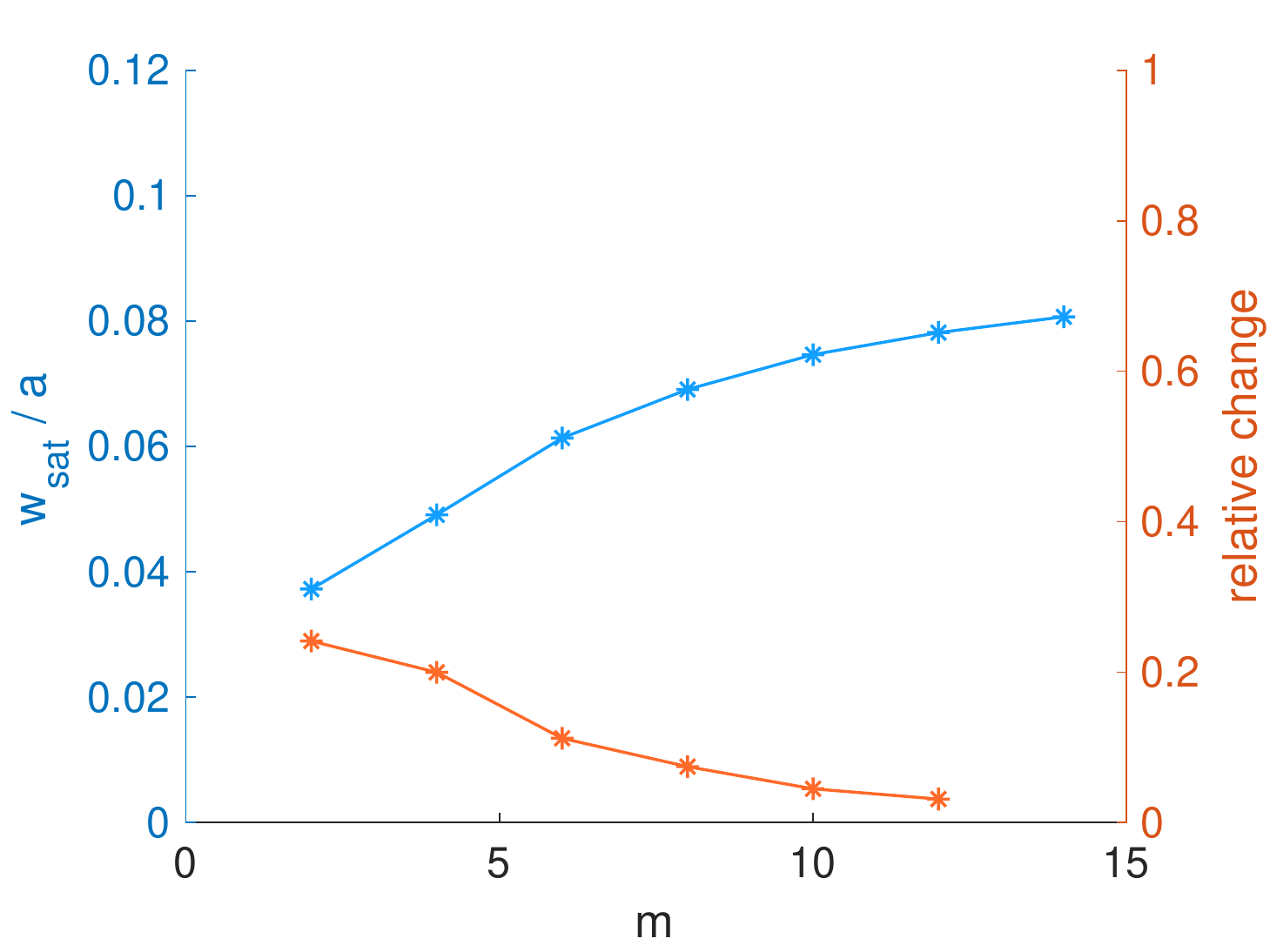}
\caption{Saturated island width (upper curve, left axis) obtained from SPEC nonlinear calculations with $q_0=1.05$ as a function of the poloidal Fourier resolution $m$. The toroidal Fourier resolution is always $n=m/2$. The relative error defined here as $(w_{sat}(m+1)-w_{sat}(m))/w_{sat}(m)$ is also shown (lower curve, right axis).} \label{fig:wconv} 
\end{figure}

Finally, we would like to remark that a convergence analysis has been carried out for the SPEC simulations presented herein. Increasing further the number $N_v=41$ of volumes does not significantly change the values of $w_{sat}$ (data not shown). Similarly, no significant changes are obtained with a further increase in the order of the radial basis functions (Tschebyshev polynomials) used to represent the magnetic vector potential in the volumes. The Fourier resolution, however, seems to play a more important role. Even if the dominant Fourier mode in the saturated island is the $(m,n)=(2,1)$ mode and higher harmonics $(m,n)=(2j,j)$, with $j\in \mathbb{N}$, are present with exponentially decreasing amplitude, the values of $w_{sat}$ can strongly depend on the Fourier resolution used in SPEC, as illustrated in Fig.~\ref{fig:wconv}. At a resolution of $(m,n)=(10,5)$, which is the one used to obtain the results in Figure \ref{fig:wsat} and already implies describing 116 different Fourier modes \cite{Hudson2020}, the relative changes in $w_{sat}$ due to a further increase in Fourier resolution become lower than $5\%$. It is also important to note that for $q_0$ close to 2, the resonant surface is close to the magnetic (and coordinate) axis, and that makes it harder for SPEC to robustly find a solution, mainly because of the singularity in the coordinates at $r=0$. Furthermore, the SPEC code is still numerically fragile in the sense that the Newton-like method employed to find force-balance might not be able to converge depending on the initial guess. The main result of this paper is thus the evidence that an equilibrium approach is capable of predicting, to a certain level of accuracy, the saturated state of tearing mode islands in a cylindrical tokamak. The potential of carrying out these predictions very fast is still to be shown by improving the numerical algorithm in SPEC. Nevertheless, when SPEC is able to find a solution, it does so in about $\sim 3$ CPU-hours, while the time-dependent resistive MHD calculations performed with SpeCyl can take hours to days, mainly depending on the value of the Lunquist number $S$, which strongly affects the simulation time required to reach saturation, and on the number of Fourier harmonics included in the calculation. The SpeCyl predictions are however more accurate, as was shown in Fig.~\ref{fig:wsat}.

\section{Conclusions} \label{sec:conc}

In this paper we have investigated the nonlinear saturation of resistive tearing modes in a zero pressure cylindrical tokamak. We have validated the results of nonlinear visco-resistive MHD simulations against an exact nonlinear saturation theory, showing excellent agreement for a wide range of equilibria. Furthermore, we have shown that an equilibrium approach can be used to predict, without free parameters, the saturated state of these modes to a certain degree of accuracy, with a very good agreement for sufficiently small islands. The natural next step is to perform similar tearing mode saturation studies in toroidal geometries. In stellarators, classical tearing modes have been observed \cite{Bartlett1980} and modeled \cite{Nikulsin2022}, and we plan on investigating them with the SPEC code. In tokamaks, neoclassical tearing modes (NTMs) play a crucial role in the potential trigger of plasma disruptions, and we plan on extending this approach with the addition of the bootstrap current effects, which are already present in the SPEC code \cite{Baillod2021, Baillod2022}. There are also toroidal corrections to classical tearing modes in tokamaks \cite{Glasser1975}, which have already been reproduced linearly with the SPEC code \cite{Kumar2023}. We plan on verifying whether the effect of these corrections on the nonlinear saturation of classical tearing modes can be retrieved with an equilibrium approach as well. Finally, the effect of rotation on tearing modes could also be studied by exploiting recent extensions of the stepped-pressure equilibrium model that include the presence of flows \cite{Qu2020}.

\section*{Acknowledgments}
The authors acknowledge useful discussions with Amitava Bhattacharjee, Susanna Cappello, Antoine Cerfon, Dominique Escande, Per Helander, Stuart Hudson, and Zhisong Qu. 

\section*{Funding}
This work has been carried out within the framework of the EUROfusion Consortium, partially funded by the European Union via the Euratom Research and Training Programme (Grant Agreement No 101052200 — EUROfusion). The Swiss contribution to this work has been funded by the Swiss State Secretariat for Education, Research and Innovation (SERI). Views and opinions expressed are however those of the author(s) only and do not necessarily reflect those of the European Union, the European Commission or SERI. Neither the European Union nor the European Commission nor SERI can be held responsible for them. This research was also supported by a grant from the Simons Foundation (1013657, JL).

\section*{Declaration of interests}
The authors report no conflict of interest.

\section*{Data availability statement}
The data that support the findings of this study is available from the authors upon reasonable request.

\section*{Author ORCID} J. Loizu, https://orcid.org/0000-0002-4862-7393; D. Bonfiglio, https://orcid.org/0000-0003-2638-317X

\section*{Appendix A}

Here we show that in cylindrical tokamak geometry, the stability of stepped-pressure equilibria can be calculated from the eigenvalues of the force-gradient matrix $\mathcal{G}$, whose coefficients measure the variation in the force on each interface, $f_l = [[B^2/2\mu_0]]_l$, with respect to perturbations in the geometry of the ideal interfaces, $r_{l,mn}$, with the constraint of fixed magnetic helicity and fluxes in each volume (Sec.~\ref{sec:specstab}). 

Consider the energy functional
 \begin{equation}
F=W-\sum_l \mu_l (K_l - K_{l,o}) \ ,
\end{equation}
where
 \begin{equation}
W = \sum_l\int_{V_l} \frac{B^2}{2\mu_0} dV \ 
\end{equation}
is the total plasma potential energy and $\mu_l$ are Lagrange multipliers defined to enforce the constraint on the magnetic helicity $K_l = K_{l,o}$ in each volume $l$. 
A variational calculation on $F$ that accounts for arbitrary variations in the magnetic field, $\delta \bv{B}=\nabla\times\delta\bv{A}$ (except for the boundary condition $\bv{B}\cdot \bv{n}=0$ and the flux constraints, which can be enforced using the gauge freedom), and interface geometry, $\delta\bv{x}\equiv \{\delta r_{l,mn}\}$, gives
 \begin{equation}
\delta F = \sum_l \int_{V_l} \left( \nabla \times \bv{B} - \mu_l \bv{B} \right)\cdot \delta \bv{A} \ dV + \sum_l \int_{\partial V_l} f_l \ \bv{\xi}_l(\delta\bv{x})\cdot d\bv{s}  \label{dF}
\end{equation}
where $\bv{\xi}_l$ is the displacement vector of the interface $l$, which depends on $\delta\bv{x}$. At equilibrium, $\delta F = 0$, $\nabla \times \bv{B} = \mu_l \bv{B}$ in each volume, and $f_l=0$ on each interface. Notice that the force $f_l$ is a function of the position on the surface describing the interface, and thus can be described, like the interface geometry, in terms of its Fourier spectrum coefficients $f_{l,mn}$, which must all be zero at equilibrium. In general, the force $\bv{f}\equiv\{f_{l,mn}\}$ is non-zero and depends on the interface geometry $\bv{x}\equiv\{ r_{l,mn} \}$, i.e. $\bv{f}=\bv{f}(\bv{x})$. We can define the force-gradient matrix as the derivative matrix $\mathcal{G}$ with coefficients
 \begin{equation}
\mathcal{G}_{ij}(\bv{x}) = \frac{\partial f_i}{\partial x_j}(\bv{x}) 
\end{equation}
and where the derivatives are taken by preserving the constraints (fluxes and helicities) fixed. The matrix $\mathcal{G}$ is square if the number of degrees of freedom (Fourier harmonics) to describe $\bv{f}$ and $\bv{x}$ are the same.

Let us now consider a perturbation from an equilibrium state $\delta\bv{x}=\bv{x}-\bv{x}_{eq}$, such that the new state satisfies a Beltrami equation with the same helicities as for the equilibrium field, but now $\bv{f}(\bv{x})\neq0$. While $\delta F = 0$ around the equilibrium point $\bv{x_{eq}}$, the variation of the energy functional around $\bv{x}$ is 
 \begin{equation}
\delta F = - \sum_l \int_{\partial V_l} f_l \ \bv{\xi}_l(\delta\bv{x})\cdot d\bv{s}  \label{dFstar}
\end{equation}
with the minus sign due to the sign convention for $\delta\bv{x}$. Now we can expand $f_l$ around $\bv{x}_{eq}$ using the fact that $f_l(\bv{x}_{eq})=0$, and write
 \begin{equation}
\delta F = - \sum_l \int_{\partial V_l} \left( \frac{\partial f_l}{\partial \bv{x}}  \biggr\rvert_{\bv{x}_{eq}} \cdot\delta\bv{x}\right) \  \bv{\xi}_l(\delta\bv{x})\cdot d\bv{s}  \quad . 
\end{equation}
Given the expansion we have performed, one finds that the infinitesimal variation of potential energy $\delta W \equiv W(\bv{x}) - W(\bv{x}_{eq})$ is thus
 \begin{equation}
\delta W = \frac{1}{2} \sum_l \int_{\partial V_l} \left( \frac{\partial f_l}{\partial \bv{x}}  \biggr\rvert_{\bv{x}_{eq}} \cdot\delta\bv{x}\right) \  \bv{\xi}_l(\delta\bv{x})\cdot d\bv{s}  \quad . \label{appdW}
\end{equation}
We now show that in cylindrical geometry, the sign of $\delta W$ in Eq.~(\ref{appdW}) is given by the sign of the eigenvalues of the matrix $\mathcal{G}$. Indeed, in cylindrical geometry, we have that the position of an interface is
\begin{equation}
\bv{r}_l(\theta,\varphi) = r_l(\theta,\varphi)\cos{\theta}\bv{\hat{i}} +   \  r_l(\theta,\varphi)\sin{\theta} \bv{\hat{j}} +  R \ \varphi \ \bv{\hat{k}}
\end{equation}
and we can expand in Fourier series the function $r_l(\theta,\varphi)$,  
\begin{equation}
r_l(\theta,\varphi) = \sum_k r_{l,k} \cos{(\alpha_k)}
\end{equation}
where $\alpha_k = m_k \theta - n_k \varphi$, with $k$ labeling all the Fourier modes in the expansion. The displacement of an interface is then given by
\begin{equation}
\bv{\xi}_l = \delta \bv{r}_l(\theta,\varphi) =  \sum_k \delta r_{l,k}\cos{(\alpha_k)} (\cos{\theta}\bv{\hat{i}} +  \sin{\theta} \bv{\hat{j}} )
\end{equation}
which expresses the relation between $\bv{\xi}_l$ and certain components of $\delta \bv{x}$. Meanwhile, the differential surface element, $d\bv{s}$, projected on the displacement vector, $\bv{\xi}_l$, gives
\begin{equation}
\bv{\xi}\cdot d\bv{s} =   R \ r_l(\theta,\varphi) \sum_k \delta r_{l,k}\cos{(\alpha_k)}   d\theta d\varphi 
\end{equation}
and we can thus express Eq.~\eqref{appdW} as
 \begin{align}
\delta W &= \frac{1}{2} \sum_l \int_{\partial V_l} \left( \frac{\partial f_l}{\partial \bv{x}}\biggr\rvert_{\bv{x}_{eq}}  \cdot\delta\bv{x}\right) \ \bv{\xi}_l(\delta\bv{x})\cdot d\bv{s}     \\
              &=\frac{1}{2}  \sum_l \int_{\partial V_l} \left( \frac{\partial f_l}{\partial \bv{x}}\biggr\rvert_{\bv{x}_{eq}}  \cdot\delta\bv{x}\right) \ R \ r_l(\theta,\varphi) \sum_k \delta r_{l,k}\cos{(\alpha_k)}   d\theta d\varphi      \\
              &=\frac{1}{2}  \sum_l\sum_k\sum_{k'} \int_{\partial V_l} \left( \frac{\partial f_{l,k'}}{\partial \bv{x}}\biggr\rvert_{\bv{x}_{eq}}  \cdot\delta\bv{x}\right)\cos{(\alpha_{k'})} \  \delta r_{l,k} \cos{(\alpha_k)}Rr_l(\theta,\varphi)d\theta d\varphi \\              
              &=\frac{R}{2}  \sum_l\sum_k\sum_{k'}  \left( \frac{\partial f_{l,k'}}{\partial \bv{x}}\biggr\rvert_{\bv{x}_{eq}}  \cdot\delta\bv{x}\right)  \delta r_{l,k}  \int_{\partial V_l} \cos{(\alpha_{k'})}  \cos{(\alpha_k)}r_l(\theta,\varphi)d\theta d\varphi \\
              &=\frac{R}{2}  \sum_l\sum_k\sum_{k'} \sum_{k''} \left( \frac{\partial f_{l,k'}}{\partial \bv{x}}\biggr\rvert_{\bv{x}_{eq}}  \cdot\delta\bv{x}\right)  \delta r_{l,k}  \ r_{l,k''}\int_{\partial V_l} \cos{(\alpha_{k'})}  \cos{(\alpha_k)} \cos{(\alpha_{k''})}d\theta d\varphi \\              
              &=\frac{R}{2}  \sum_l\sum_k\sum_{k'} \sum_{k''} \left( \frac{\partial f_{l,k'}}{\partial \bv{x}}\biggr\rvert_{\bv{x}_{eq}}  \cdot\delta\bv{x}\right)  \delta r_{l,k}  \ r_{l,k''} \mathcal{I}_{kk'k''}
\end{align}
where 
\begin{equation}
\mathcal{I}_{kk'k''} = \oint\oint \cos{(\alpha_{k})}  \cos{(\alpha_{k'})} \cos{(\alpha_{k''})}d\theta d\varphi  \ .
\end{equation}

Finally, assuming that the equilibrium is axisymmetric, $r_{l,k}=r_{l,0}\delta_{k,0}$, and that $\delta\bv{x}$ is an eigenvector of $\mathcal{G}$ with eigenvalue $\lambda$, $\mathcal{G}\cdot\delta\bv{x}=\lambda \delta\bv{x}$, we have
 \begin{align}
\delta W &= \frac{R}{2} \sum_l\sum_k\sum_{k'}  \left( \frac{\partial f_{l,k'}}{\partial \bv{x}}\biggr\rvert_{\bv{x}_{eq}}  \cdot\delta\bv{x}\right)  \delta r_{l,k}  \ r_{l,0} \mathcal{I}_{kk'0}    \\           
              &= \lambda\frac{R}{2} \sum_l\sum_k\sum_{k'}  \delta r_{l,k'}   \delta r_{l,k}  \ r_{l,0} \mathcal{I}_{kk'0} \\
              &= \lambda\pi^2R\sum_l\sum_k\sum_{k'}  \delta r_{l,k'}   \delta r_{l,k}  \ r_{l,0} \delta_{k,k'} 	\\
              &= \lambda\pi^2R\sum_l\sum_k  \delta r_{l,k}^2  \ r_{l,0} 
\end{align}
and hence the sign of $\delta W$ is given by the sign of $\lambda$. Hence the stability of an axisymmetric equilibrium can therefore be inferred from the (sign of the) eigenvalues of $\mathcal{G}$.

\section*{References}

\bibliography{spc_bibliography}{}

\begin{thebibliography}{10}

\bibitem{White2014}
Roscoe~B White.
\newblock {\em {The Theory of Toroidally Confined Plasmas}}.
\newblock IMPERIAL COLLEGE PRESS, jan 2014.

\bibitem{LaHaye2006}
R.~J. {La Haye}.
\newblock {Neoclassical tearing modes and their control}.
\newblock {\em Physics of Plasmas}, 13(5):055501, may 2006.

\bibitem{Poye2014}
A.~Poy{\'{e}}, O.~Agullo, A.~Smolyakov, S.~Benkadda, and X.~Garbet.
\newblock {Dynamics of magnetic islands in large $\Delta^\prime$ regimes}.
\newblock {\em Physics of Plasmas}, 21(2), 2014.

\bibitem{Cooper2010}
W.~A. Cooper, J.~P. Graves, A.~Pochelon, O.~Sauter, and L.~Villard.
\newblock {Tokamak magnetohydrodynamic equilibrium states with axisymmetric
  boundary and a 3D helical core}.
\newblock {\em Physical Review Letters}, 105(3):3--6, 2010.

\bibitem{Cooper2011}
W~A Cooper, J~P Graves, O~Sauter, I~T Chapman, M~Gobbin, L~Marrelli, P~Martin,
  I~Predebon, and D~Terranova.
\newblock {Magnetohydrodynamic properties of nominally axisymmetric systems
  with 3D helical core}.
\newblock {\em Plasma Physics and Controlled Fusion}, 53(7):074008, 2011.

\bibitem{Kleiner2019}
A~Kleiner, J~P Graves, D~Brunetti, W~A Cooper, S~Medvedev, A~Merle, and
  C~Wahlberg.
\newblock {Current and pressure gradient triggering and nonlinear saturation of
  low- n edge harmonic oscillations in tokamaks}.
\newblock {\em Plasma Physics and Controlled Fusion}, 61(8):084005, aug 2019.

\bibitem{Dennis2013a}
G.~R. Dennis, S.~R. Hudson, D.~Terranova, P.~Franz, R.~L. Dewar, and M.~J.
  Hole.
\newblock {Minimally constrained model of self-organized helical states in
  reversed-field pinches}.
\newblock {\em Physical Review Letters}, 111(5):1--5, 2013.

\bibitem{Loizu2020}
J.~Loizu, Y.-M. Huang, S.R. Hudson, A.~Baillod, A.~Kumar, and Z.S. Qu.
\newblock {Direct prediction of nonlinear tearing mode saturation using a
  variational principle}.
\newblock {\em Physics of Plasmas}, 27(7), 2020.

\bibitem{Wright2022}
A.M. Wright, P~Kim, N.M. Ferraro, and S.R. Hudson.
\newblock {Modelling the nonlinear plasma response to externally applied
  three-dimensional fields with the Stepped Pressure Equilibrium Code}.
\newblock {\em Journal of Plasma Physics}, 88(5):905880508, oct 2022.

\bibitem{Arcis2006}
N.~Arcis, D.~F. Escande, and M.~Ottaviani.
\newblock {Rigorous approach to the nonlinear saturation of the tearing mode in
  cylindrical and slab geometry}.
\newblock {\em Physics of Plasmas}, 13(5):052305, may 2006.

\bibitem{Cappello1996}
S.~Cappello and D.~Biskamp.
\newblock {Reconnection processes and scaling laws in reversed field pinch
  magnetohydrodynamics}.
\newblock {\em Nuclear Fusion}, 36(5):571--581, 1996.

\bibitem{Bonfiglio2010}
D.~Bonfiglio, L.~Chac{\'{o}}n, and S.~Cappello.
\newblock {Nonlinear three-dimensional verification of the SPECYL and PIXIE3D
  magnetohydrodynamics codes for fusion plasmas}.
\newblock {\em Physics of Plasmas}, 17(8), 2010.

\bibitem{Hudson2012}
S.~R. Hudson, R.~L. Dewar, G.~Dennis, M.~J. Hole, M.~McGann, G.~{Von Nessi},
  and S.~Lazerson.
\newblock {Computation of multi-region relaxed magnetohydrodynamic equilibria}.
\newblock {\em Physics of Plasmas}, 19(11), 2012.

\bibitem{Hudson2020}
S.~R. Hudson, J.~Loizu, C.~Zhu, Z.~S. Qu, C.~N{\"{u}}hrenberg, S.~Lazerson,
  C.~B. Smiet, and M.~J. Hole.
\newblock {Free-boundary MRxMHD equilibrium calculations using the
  stepped-pressure equilibrium code}.
\newblock {\em Plasma Physics and Controlled Fusion}, 62(8), 2020.

\bibitem{Freidberg2014}
J~P Freidberg.
\newblock {\em {ideal MHD}}.
\newblock Cambridge University Press, 2014.

\bibitem{Furth1973}
H.~P. Furth, P.~H. Rutherford, and H.~Selberg.
\newblock {Tearing mode in the cylindrical tokamak}.
\newblock {\em Physics of Fluids}, 16(7):1054--1063, 1973.

\bibitem{Dennis2013}
G.~R. Dennis, S.~R. Hudson, R.~L. Dewar, and M.~J. Hole.
\newblock {The infinite interface limit of multiple-region relaxed
  magnetohydrodynamics}.
\newblock {\em Physics of Plasmas}, 20(3), 2013.

\bibitem{Baillod2021}
A.~Baillod, J.~Loizu, Z.S. Qu, A.~Kumar, and J.P. Graves.
\newblock {Computation of multi-region, relaxed magnetohydrodynamic equilibria
  with prescribed toroidal current profile}.
\newblock {\em Journal of Plasma Physics}, 87(4), 2021.

\bibitem{Loizu2019a}
J.~Loizu and S.R. Hudson.
\newblock {Multi-region relaxed magnetohydrodynamic stability of a current
  sheet}.
\newblock {\em Physics of Plasmas}, 26(3), 2019.

\bibitem{Kumar2021}
A.~Kumar, Z.~Qu, M.~J. Hole, A.~M. Wright, J.~Loizu, S.~R. Hudson, A.~Baillod,
  R.~L. Dewar, and N.~M. Ferraro.
\newblock {Computation of linear MHD instabilities with the multi-region
  relaxed MHD energy principle}.
\newblock {\em Plasma Physics and Controlled Fusion}, 63(4), 2021.

\bibitem{Kumar2022}
A.~Kumar, C.~N{\"{u}}hrenberg, Z.~Qu, M.~J. Hole, J.~Doak, R.~L. Dewar, S.~R.
  Hudson, J.~Loizu, K.~Aleynikova, A.~Baillod, and H.~Hezaveh.
\newblock {Nature of ideal MHD instabilities as described by multi-region
  relaxed MHD}.
\newblock {\em Plasma Physics and Controlled Fusion}, 64(6), 2022.

\bibitem{Rutherford1973}
P.~H. Rutherford.
\newblock {Nonlinear growth of the tearing mode}.
\newblock {\em Physics of Fluids (1958-1988)}, 16(11):1903--1908, 1973.

\bibitem{Cappello2004}
S.~Cappello.
\newblock {Bifurcation in the MHD behaviour of a self-organizing system: The
  reversed field pinch (RFP)}.
\newblock {\em Plasma Physics and Controlled Fusion}, 46(12 B), 2004.

\bibitem{Teng2016}
Q.~Teng, N.~Ferraro, D.~A. Gates, S.~C. Jardin, and R.~B. White.
\newblock {Nonlinear asymmetric tearing mode evolution in cylindrical
  geometry}.
\newblock {\em Physics of Plasmas}, 23(10), 2016.

\bibitem{White1977}
Roscoe~B White, D~A Monticello, Marshall~N Rosenbluth, and B~V Waddell.
\newblock {Saturation of the tearing mode}.
\newblock {\em Physics of Fluids}, 20, 1977.

\bibitem{Fitzpatrick1995}
R~Fitzpatrick.
\newblock {Helical temperature perturbations associated with radially
  asymmetric magnetic island chains in tokamak plasmas}.
\newblock {\em Physics of Plasmas}, 23(12), 2016.

\bibitem{Wesson2004}
J~Wesson.
\newblock {\em {Tokamaks}}.
\newblock Clarendon Press, Oxford, 2004.

\bibitem{Berger1999}
Mitchell~A Berger.
\newblock {Introduction to magnetic helicity}.
\newblock {\em Plasma Physics and Controlled Fusion}, 41:166--176, 1999.

\bibitem{Heidbrink2000}
W.W. Heidbrink and T.H. Dang.
\newblock {Magnetic helicity is conserved at a tokamak sawtooth crash}.
\newblock {\em Plasma Physics and Controlled Fusion}, 42:L31--L36, 2000.

\bibitem{Militello2004}
F~Militello and F~Porcelli.
\newblock {Simple analysis of the nonlinear saturation of the tearing mode}.
\newblock {\em Physics of Plasmas}, 11, 2004.

\bibitem{Escande2004}
D~F Escande and M~Ottaviani.
\newblock {Simple and rigorous solution for the nonlinear tearing mode}.
\newblock {\em Physics Letters, Section A: General, Atomic and Solid State
  Physics}, 323:278--284, 2004.

\bibitem{Bartlett1980}
D.~V. Bartlett, G.~Cannici, G.~Cattanei, D.~Dorst, A.~Eisner, G.~Grieger,
  H.~Hacker, J.~How, H.~J{\"{a}}ckel, R.~Jaenicke, P.~Javel, J.~Junker,
  M.~Kick, R.~Lathe, F.~Leuterer, C.~Mahn, S.~Marlier, G.~M{\"{u}}ller,
  W.~Ohlendorf, F.~Rau, H.~Renner, H.~Ringler, J.~Saffert, J.~Sapper,
  P.~Smeulders, M.~Tutter, A.~Weller, E.~W{\"{u}}rsening, and H.~Wobig.
\newblock {Stabilization of the (2, 1) tearing mode and of the current
  disruption in the W VII–A stellarator}.
\newblock {\em Nuclear Fusion}, 20(9):1093--1100, sep 1980.

\bibitem{Nikulsin2022}
N.~Nikulsin, R.~Ramasamy, M.~Hoelzl, F.~Hindenlang, E.~Strumberger, K.~Lackner,
  and S.~G{\"{u}}nter.
\newblock {JOREK3D: An extension of the JOREK nonlinear MHD code to
  stellarators}.
\newblock {\em Physics of Plasmas}, 29(6):063901, jun 2022.

\bibitem{Baillod2022}
A.~Baillod, J.~Loizu, Z.~Qu, H.~P. Arbez, and J.~P. Graves.
\newblock {Equilibrium beta-limits dependence on bootstrap current in classical
  stellarators}.
\newblock {\em submitted to Journal of Plasma Physics}, pages 1--20, nov 2022.

\bibitem{Glasser1975}
A.~H Glasser, J.~M Greene, and J.~L Johnson.
\newblock {Resistive instabilities in general toroidal plasma configurations}.
\newblock {\em Physics of Fluids}, 18(7):875, 1975.

\bibitem{Kumar2023}
Arunav Kumar, Joaquim Loizu, Matthew~J Hole, Zhisong Qu, Stuart~R Hudson, and
  Robert~L Dewar.
\newblock {On the relationship between the multi-region relaxed variational
  principle and resistive inner layer theory}.
\newblock {\em Plasma Physics and Controlled Fusion}, pages 0--29, mar 2023.

\bibitem{Qu2020}
Z.~S. Qu, R.~L. Dewar, F.~Ebrahimi, J.~K. Anderson, S.~R. Hudson, and M.~J.
  Hole.
\newblock {Stepped pressure equilibrium with relaxed flow and applications in
  reversed-field pinch plasmas}.
\newblock {\em Plasma Physics and Controlled Fusion}, 62(5):054002, may 2020.

\end{thebibliography}
\bibliographystyle{unsrt}

\newpage

%
%

\end{document}